\documentclass[lettersize,journal]{IEEEtran}

\ifCLASSOPTIONcompsoc
  \usepackage[nocompress]{cite}
\else
  \usepackage{cite}
\fi

\ifCLASSINFOpdf
\else
\fi

\usepackage[pdftex]{hyperref}
\usepackage{pifont}
\usepackage[ruled]{algorithm2e}
\usepackage{my-tvcg}
\usepackage{epigraph}
\usepackage{float}
\usepackage{diagbox}
\usepackage{multirow}
\usepackage{color}
\usepackage{placeins}

\graphicspath{{./images/}}

\definecolor{MainColor}{RGB}{128, 0, 128}

\hypersetup{
    unicode=true,
    pdftoolbar=true,
    pdfmenubar=true,
    pdffitwindow=false,
    pdfstartview={FitH},
    pdftitle={Efficient Computation of Integer-constrained Cones for Conformal Parameterizations},
    pdfnewwindow=true,
    colorlinks=true,
    linkcolor=MainColor,
    citecolor=MainColor,
    filecolor=magenta,
    urlcolor=MainColor,
    breaklinks=MainColor
}

\definecolor{curvered}{RGB}{255,110,107}
\definecolor{curveblue}{RGB}{128, 160, 255}
\definecolor{curveorange}{RGB}{250, 152, 58}
\definecolor{curvegreen}{RGB}{111, 219, 133}
\definecolor{notescolor}{RGB}{222,127,177}

\newcommand{\textblue}{\textcolor[rgb]{0.0,0.0,0.0}}
\definecolor{color1}{rgb}{0.51, 0.43, 0.78}
\definecolor{color2}{rgb}{0.55, 0.824, 0.706}

\begin{document}

\title{Efficient Computation of Integer-constrained Cones for Conformal Parameterizations}

\author{Wei~Du, Qing~Fang, Ligang~Liu, Xiao-Ming~Fu
\IEEEcompsocitemizethanks
{
\IEEEcompsocthanksitem W. Du, Q. Fang, L. Liu, and X. Fu are with the School of Mathematical Sciences, University of Science and Technology of China. 
\protect\\
E-mail: duweiyou@mail.ustc.edu.cn, fq1208@mail.ustc.edu.cn, lgliu@ustc.edu.cn, fuxm@ustc.edu.cn.
\IEEEcompsocthanksitem Corresponding author: Qing~Fang.

Under review at IEEE Transactions on Visualization and Computer Graphics (TVCG). This work has been submitted to the IEEE for possible publication. Copyright may be transferred without notice, after which this version may no longer be accessible.
}
}

\markboth{Under review}
{Shell \MakeLowercase{\textit{et al.}}: Efficient Computation of Integer-constrained Cones for Conformal Parameterizations}

\IEEEtitleabstractindextext{%
\begin{abstract}
We propose an efficient method to compute a small set of integer-constrained cone singularities, which induce a rotationally seamless conformal parameterization with low distortion.
Since the problem only involves discrete variables, i.e., vertex-constrained positions, integer-constrained angles, and the number of cones, we alternately optimize these three types of variables to achieve tractable convergence.
Central to high efficiency is an explicit construction algorithm that reduces the optimization problem scale to be slightly greater than the number of integer variables for determining the optimal angles with fixed positions and numbers, even for high-genus surfaces. 
In addition, we derive a new derivative formula that allows us to move the cones, effectively reducing distortion until convergence.
Combined with other strategies, including repositioning and adding cones to decrease distortion, adaptively selecting a constrained number of integer variables for efficient optimization, and pairing cones to reduce the number, we quickly achieve a favorable tradeoff between the number of cones and the parameterization distortion.
We demonstrate the effectiveness and practicability of our cones by using them to generate rotationally seamless and low-distortion parameterizations on a massive test data set.
Our method demonstrates an order-of-magnitude speedup (30$\times$ faster on average) compared to state-of-the-art approaches while maintaining comparable cone numbers and parameterization distortion.
\end{abstract}

}

\maketitle
\IEEEdisplaynontitleabstractindextext
\IEEEpeerreviewmaketitle

\section{Introduction} \label{sec:intro}
Conformal parameterization, a key technique in computer graphics and geometric processing, plays a vital role in various applications, such as texture mapping~\cite{Desbrun2002Intrinsic,levy2002least}, surface remeshing~\cite{Alliez2003,Zhong2014}, and physical simulation~\cite{Segall2016}, primarily due to its angle-preserving property. 
However, conformal parameterization typically suffers from significant area distortion, which can severely compromise the accuracy of downstream applications. 
Introducing cone singularities has proven effective in mitigating such distortion~\cite{Springborn2008,myles2012global,kharevych2006discrete,ben2008conformal,fang2021computing,Soliman2018OCS,Li2022int,zhang2023practical}, forming cone parameterizations successfully applied to many applications, including 4D printing~\cite{Nojoomi2021}, developable approximation~\cite{Zhao2023}, and anisotropic remeshing~\cite{dai2024anisotropic}.

\begin{figure}[t]
  \centering
  \vspace{-1mm}
  \begin{overpic}[width=0.99\linewidth]{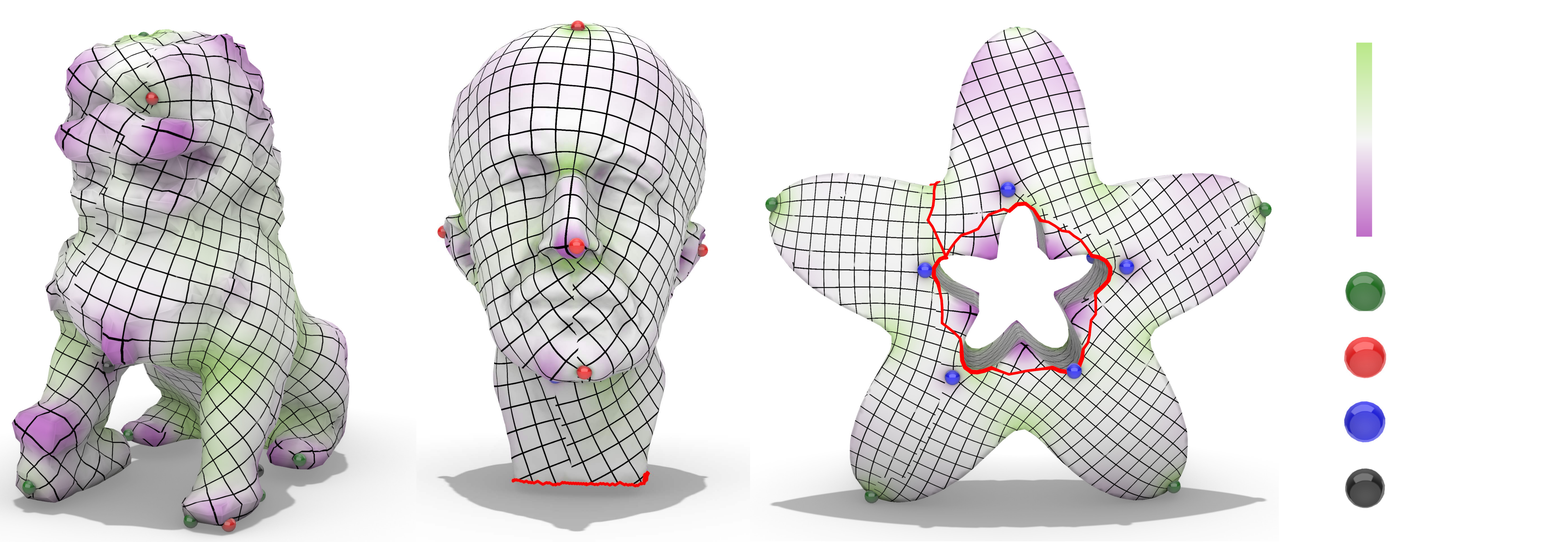}
    {
 \put(90.5,31){\small \(0.5\)}
 \put(89.5,19){\small \(-0.5\)}
 \put(90,15){\small \(\ge \pi\)}
 \put(90,11){\small \(=\frac{\pi}{2}\)}
 \put(90,6.7){\small \(=-\frac{\pi}{2}\)}
 \put(90,2.6){\small \(\le -\pi\)}

 \put(1,-3){\small (17, 0.196, 0.9)}
 \put(26,-3){\small (10, 0.119, 18.5)}
 \put(54,-3){\small (15, 0.145, 2.0)}
    }
  \end{overpic}
  \vspace{-1.5mm}
  \caption{
  Integer-constrained cone singularities for genus-zero meshes without boundaries (left, \(N=4k\)), meshes with boundaries (middle, \(N=50k\)), and nonzero genus meshes (right, \(N=10k\)), where \(N\) denotes the number of mesh vertices.
  The triplet below each model shows the cone number, the distortion (defined in Sec.~\ref{subsec:problem_definition}), and the runtime (in seconds). All subsequent figures use this triplet representation unless noted.
  Discrete log conformal factors via color bar and cones via colored spheres are visualized.
  }
  \label{fig:teaser}
\end{figure}
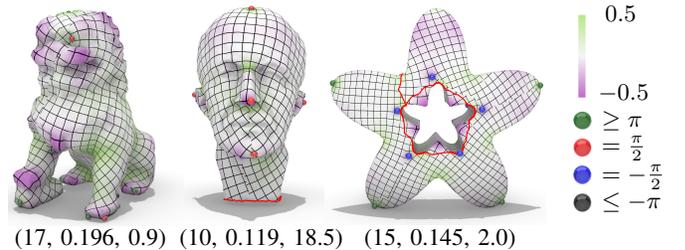

Computing cone singularities involves determining their \emph{positions, angles, and number} to reduce parameterization distortion (Fig.~\ref{fig:teaser}). 
These singularities are typically constrained to mesh vertices on the input triangular mesh, making their positions discrete variables.
To construct global seamless parameterizations, which can be obtained through rotationally seamless parameterizations~\cite{campen2021efficient,Campen2019Seamless,myles2012global,campen2017similarity}, the cone angles and holonomy angles for the non-contractible homology loops must strictly satisfy the constraint of being an integer multiple of $\pi/2$. 
Furthermore, it is generally desirable to minimize the number of singularities. 
Consequently, the discrete nature of all variables introduces significant challenges in achieving a favorable tradeoff among low parameterization distortion, small number of cones, and high efficiency.

\begin{figure}[t]
  \centering
  \begin{overpic}[width=0.9\linewidth]{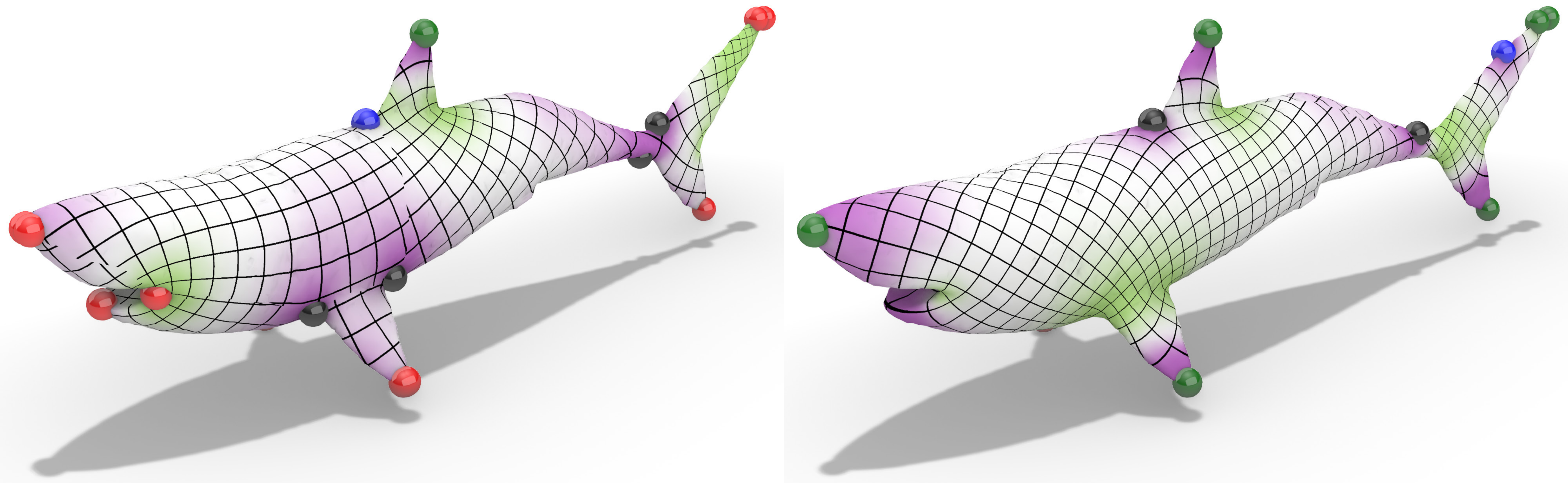}
    {
 \put(11,-3){\small (25, 0.237, 251.0)}
 \put(40,3){\small \cite{Li2022int}}
 \put(64,-3){\small  (15, 0.230, 3.7)}
 \put(90,3){\small Ours}
    }
  \end{overpic}
  \vspace{3mm}
  \caption{
  Comparison with~\cite{Li2022int}. 
  We achieve fewer cones, lower distortion, and a much shorter time (in seconds, 60$\times$ faster). 
  }
  \label{fig:cmp-Li}
\end{figure}

\begin{figure}[t]
  \centering
  \begin{overpic}[width=0.8\linewidth]{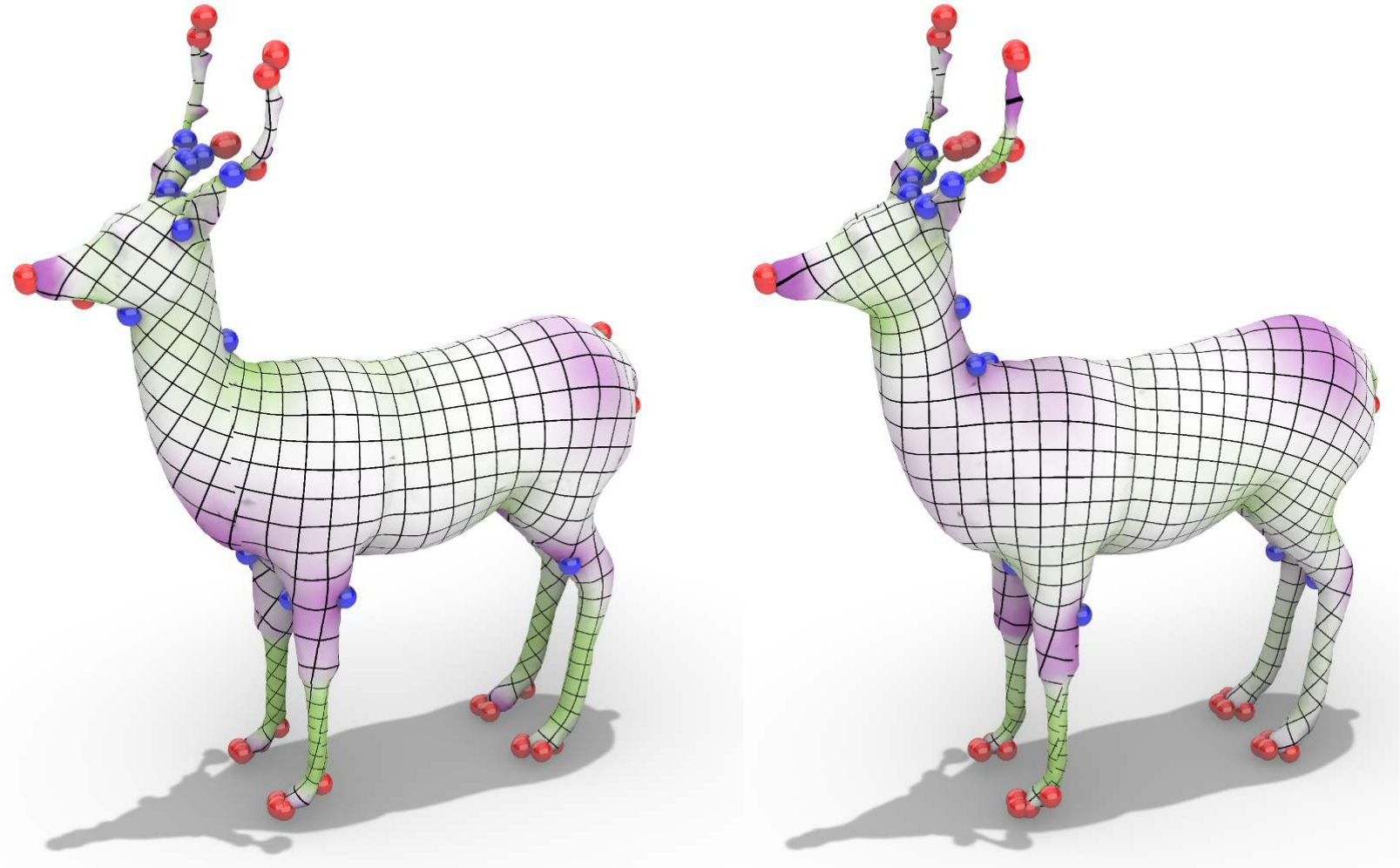}
    {
 \put(10,-3){\small(60, 0.244, 1804.3)}
 \put(40,3){\small \cite{zhang2023practical}}
 \put(64,-3){\small(58, 0.237, 9.9)}
 \put(90,3){\small Ours}
    }
  \end{overpic}
  \vspace{3mm}
  \caption{
  Comparison with~\cite{zhang2023practical}. Fewer cones and lower distortion are obtained by our method in much less time (180$\times$ faster).
  }
  \label{fig:cmp-zhang}
\end{figure}
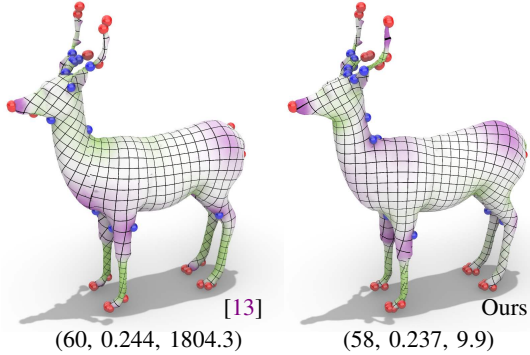

Several methods have been proposed for this challenging problem~\cite{myles2012global,Li2022int,zhang2023practical}. 
\cite{myles2012global} develop a progressive curvature concentration strategy to determine the positions and number of cone singularities, while constraining their angles to integer multiples of $\pi/2$ through iterative rounding. However, the two progressive mechanisms often produce suboptimal results with excessive cone points and parameterization distortion.
\cite{Li2022int} formulate the problem as a mixed-integer optimization solved via the Douglas-Rachford splitting algorithm. Since the number of integer-constrained variables equals the input vertex number, it suffers from high computational costs and a vast search space that frequently causes the optimization to converge to local optima, ultimately failing to achieve a favorable balance between competing objectives (Fig.~\ref{fig:cmp-Li}).
\cite{zhang2023practical} rely on the enumeration of angles and positions to co-optimize singularity count and parameterization distortion, demonstrating improved tradeoff capability. However, this exhaustive enumeration process dramatically increases computational complexity (Fig.~\ref{fig:cmp-zhang}).

This paper proposes an efficient method for generating sparse cone singularities with integer constraints, enabling the construction of rotationally seamless conformal parameterizations with low distortion.
\textblue{Our method is inspired by~\cite{li2023efficient}, which adopts an alternating optimization approach for position, angle, and number of cones between three steps:
\begin{itemize}
    \item Moving cones along the distortion energy gradient.
    \item Optimizing cone angles to reduce distortion.
    \item Inserting cones using a greedy branch-driven strategy.
\end{itemize}}

\textblue{However,~\cite{li2023efficient} only works effectively when cone angles are unconstrained. Adding integer constraints into their framework causes two major challenges:
\begin{itemize}
    \item Solving a large-scale MIQP (with many continuous and integer variables) during cone angle optimization becomes impractical.
    \item Optimizing cone positions with integer-constrained angles fails to decrease the distortion using their strategy.
\end{itemize}}

\textblue{We make two key contributions to improve efficiency:
\begin{itemize}
\item \emph{Efficient cone angle optimization.} We formulate an MIQP for cone angle optimization and exploit structural properties of the constraint matrix for variable reduction, handling both genus-zero and high-genus cases, while further limiting variables by selecting a subset of cones.
\item \emph{Alternating optimization with fixed angles.} We alternate between moving cones with fixed angles and re-optimizing angles. Moreover, we derive a new gradient expression to effectively decrease distortion.
\end{itemize}}

Our algorithm efficiently balances sparse integer-constrained cone placement with low distortion.
We evaluate it on a dataset comprising 3885 models to demonstrate its capability and practicability.
Compared with the state-of-the-art methods~\cite{Li2022int,zhang2023practical}, we achieve an order of magnitude speedup (30 times faster on average) for comparable cone density and distortion.

\section{Related work} \label{sec:related}
\paragraph{Cone singularities}
\cite{kharevych2006discrete} first introduce cone singularities for conformal parameterizations. 
Based on their Gaussian curvature distribution, cone singularities can be classified into two types: cones with arbitrary angles and quantized cones. 
Cones with arbitrary angles are primarily employed to reduce area distortion in conformal parameterizations.
Typically, cones are iteratively inserted at vertices exhibiting high distortion, and the log conformal factors are computed either by solving a convex optimization problem~\cite{Springborn2008} or through a diffusion process~\cite{ben2008conformal}.
\cite{myles2012global} propose an iterative approach that progressively flattens regions with low Gaussian curvature, ultimately identifying cone singularities at isolated vertices with high curvature. 
By formulating an objective that balances area distortion and cone sparsity, global optimization methods have been developed to determine the optimal placement of cones, including convex $\ell_1$ optimization~\cite{Soliman2018OCS} and non-convex $\ell_0$ optimization~\cite{fang2021computing}.
\cite{li2023efficient} enhance efficiency by iteratively relocating cones using shape derivatives and introducing new cones to reduce parameterization distortion.
Quantized cones are necessary for computing globally or rotationally seamless parameterizations, where cone angles are constrained to be integer multiples of $\frac{\pi}{2}$.
\cite{myles2012global} progressively round the unconstrained cone angles to the nearest multiples of $\frac{\pi}{2}$, which unavoidably leads to increased area distortion.
To better control distortion, the quantized cone generation is instead formulated as a mixed-integer optimization problem, which can be solved using the Douglas-Rachford scheme~\cite{Li2022int} or a series of second-order cone programming steps~\cite{zhang2023practical}.
However, solving mixed-integer problems at scales comparable to the input mesh vertex number is expensive. 
%
Instead, we propose several explicit constructions that dramatically reduce the problem scale for high computational efficiency.

\paragraph{Field singularities}
Recent surveys~\cite{de2016vector,vaxman2016} on vector field design review various approaches for determining the number, indices, and placement of singularities that can also be used for conformal parameterizations.
On discrete triangular meshes, vector fields can be represented in several ways, including face-based~\cite{polthier2003identifying,tong2003discrete}, edge-based~\cite{fisher2007design,wang2006edge}, and vertex-based~\cite{zhang2006vector,knoppel2013globally,liu2016discrete} formulations.
Various discretizations have been proposed, such as angle-based~\cite{bommes2009mixed,farchi2018integer}, complex-number~\cite{knoppel2013globally}, complex polynomials~\cite{diamanti2014designing}, and discrete 1-forms~\cite{fisher2007design,ben2010discrete,wang2006edge} representations.
These methods define a smoothness energy optimized to determine vector fields and singularities, grounded in discrete analogies~\cite{crane2010trivial,crane2013digital} to parallel transport and trivial connections on smooth manifolds. 
Since the parameterization is often generated by tracing vector fields, no explicit correlation exists between parameterization distortion and singularity number.
\cite{coiffier2023method} employ Cartan’s method of moving frames to relate parametrization to field singularities, formulating a nonlinear least-squares problem that simultaneously places singularities and minimizes distortion.
Although it allows for penalizing non-conformal distortion, the large number of frame variables and multiple constraints make the algorithm inefficient.

\paragraph{Integer-constrained optimization in geometric processing}
Integer-constrained optimization is common in geometric processing applications, e.g., vector field design, cone generation, and quad meshing~\cite{vaxman2016,Bommes2013}. 
However, the integer-constrained optimization problem is generally an NP-hard problem~\cite{gorry1972relaxation,floudas1995nonlinear}.
Therefore, rounding methods are mainly adopted in these methods.
The direct method performs rounding on all variables simultaneously~\cite{kalberer2007quadcover}.
The iterative approach incrementally rounds one continuous variable to an integer at a time, updating the remaining variables after each step~\cite{bommes2009mixed,myles2012global}.
Two-stage methods first reduce the problem size through an approximate equivalence transformation, and then apply branch-and-bound rounding to obtain an approximate solution~\cite{bommes2013integer}.
Unlike these rounding-based strategies, the approach in~\cite{farchi2018integer} begins with a feasible solution and iteratively improves it using descent steps that strictly maintain integer constraints.
\cite{Li2022int} convert the integer constraints to binary constraints and then replace them with the intersection of a box set and a sphere to solve it in a continuous domain.
\cite{zhang2023practical} use the second-order cone programming to approximate and solve integer-constrained optimization.
We develop a reduction framework to solve the integer-constrained optimization problem with high efficiency.
\section{Method for Genus-zero Surfaces} \label{sec:method}

\subsection{Problem and methodology}\label{subsec:problem_definition}

\paragraph{Discrete conformal parameterizations}
For a given Riemannian manifold, the change of Gaussian curvature under a conformal mapping can be described with the Yamabe equation~\cite{aubin2013some}. 
We discuss the case of surfaces without boundaries. For surfaces with boundaries, we defer it to \textblue{supplements}.
For genus-zero triangular mesh \(\mathcal{M}\) with \(N\) vertices without boundaries, given the positive-semidefinite discrete Laplacian matrix \(\mathbf{L}\) using cotangent weights, the FEM approximation of the Yamabe equation is~\cite{aubin2013some,ben2008conformal,kharevych2006discrete,Springborn2008}
\begin{equation}
\mathbf{L}\mathbf{u}=\mathbf{k}^\text{tar}-\mathbf{k}^\text{ori},
\end{equation}
where \(\mathbf{u},\mathbf{k}^\text{tar}\) and \(\mathbf{k}^\text{ori}\) are \(N\)-dimensional vectors and their \(i\)-th entries \(u_i,k_i^\text{tar}\) and \(k_i^\text{ori}\) represent the discrete log conformal factor, the target Gaussian curvature, and the original Gaussian curvature at each vertex \(v_i\), respectively.

\paragraph{Area distortion}\label{para:area_distortion}
Since the Hencky energy~\cite{sharp2018variational} remains bounded for any solution to the Yamabe equation~\cite{Soliman2018OCS}, we adopt it as the metric to quantify the conformal parameterization distortion, which is defined as 
    $\mathcal{E} = \left( \int_{\mathcal{M}}u^2dA \right)^{\frac{1}{2}}.$
This \(l_2\)-norm form can be discretized as \(\mathcal{E}(\mathbf{u})=\sqrt{\mathbf{u}^T\mathbf{A}\mathbf{u}}\), where \(\mathbf{A}\) is \tb{an area weighted} matrix and its \(i\)-th diagonal element \(\mathbf{A}_{ii}\) is computed as \(\mathbf{A}_{ii}=\frac{\text{Area}(v_i)}{\sum_{i=1}^{N}\text{Area}(v_i)}\), where \(\text{Area}(v_i)=\sum_{t\in \Omega_i}\text{Area}(t)/3\) and \(\Omega_i\) is the set of triangles incident to \(v_i\).

\paragraph{\textblue{The problem} of optimizing cone configuration}\label{para:optimizing_cone_configuration}
As a special kind of conformal mapping, cone parameterizations require \(\mathbf{k}^\text{tar}\) to be zero except at finite cone singularities~\cite{BUNIN2008MeshGen}. 
Similar to~\cite{Li2022int,zhang2023practical,myles2012global}, we aim to find $N_c$ vertices \(\mathcal{C}=\{c_i\}_{i=1}^{N_c}\) as cone singularities for conformal parameterization under the following constraints and objectives.  
\begin{itemize}
    \item \emph{Topology constraints}: the target curvature obeys the Gauss-Bonnet theorem, i.e., the sum of entries of \(\mathbf{k}^\text{tar}\) is \(4\pi\).
    \item \emph{Integer constraints}: the parameterization is global seamless, requiring each entry of \(\mathbf{k}^\text{tar}\) to be an integer multiple of \(\pi/2\).
    \item \emph{Low distortion objectives}: the parameterization distortion \(\mathcal{E}\) is smaller than a specified threshold $\epsilon_\text{tar}$.
    \item  \emph{Cone efficiency objectives}: \tb{introduce necessary cones to reduce distortion and avoid redundant ones.}
\end{itemize}



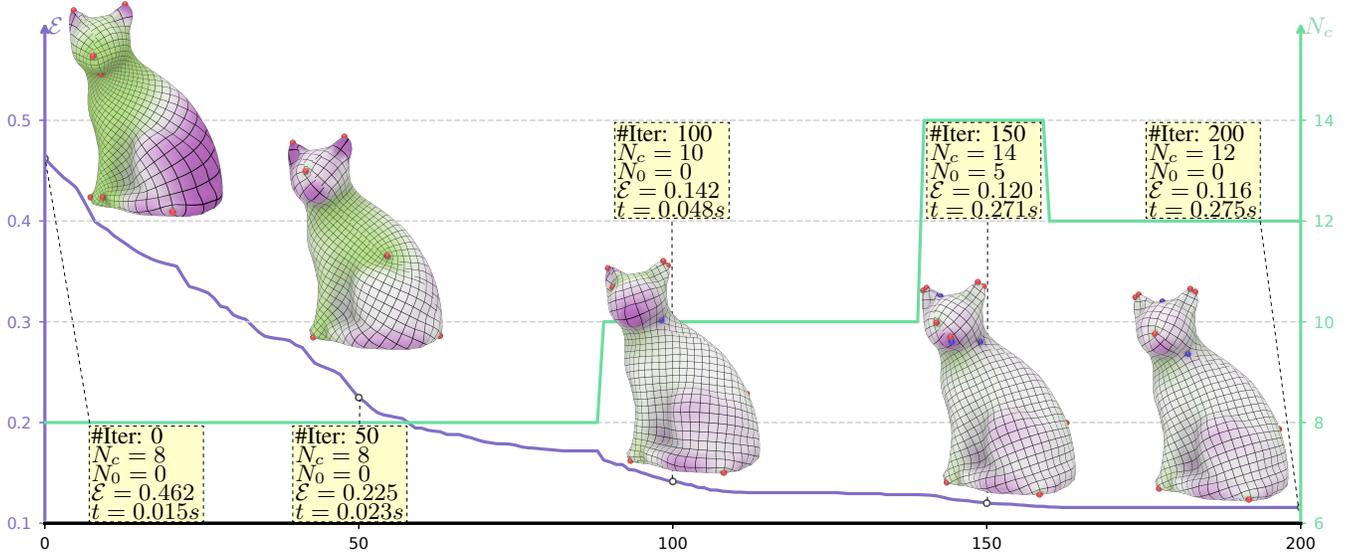
\begin{figure*}[t]
  \centering
  \begin{overpic}[width=0.99\linewidth]{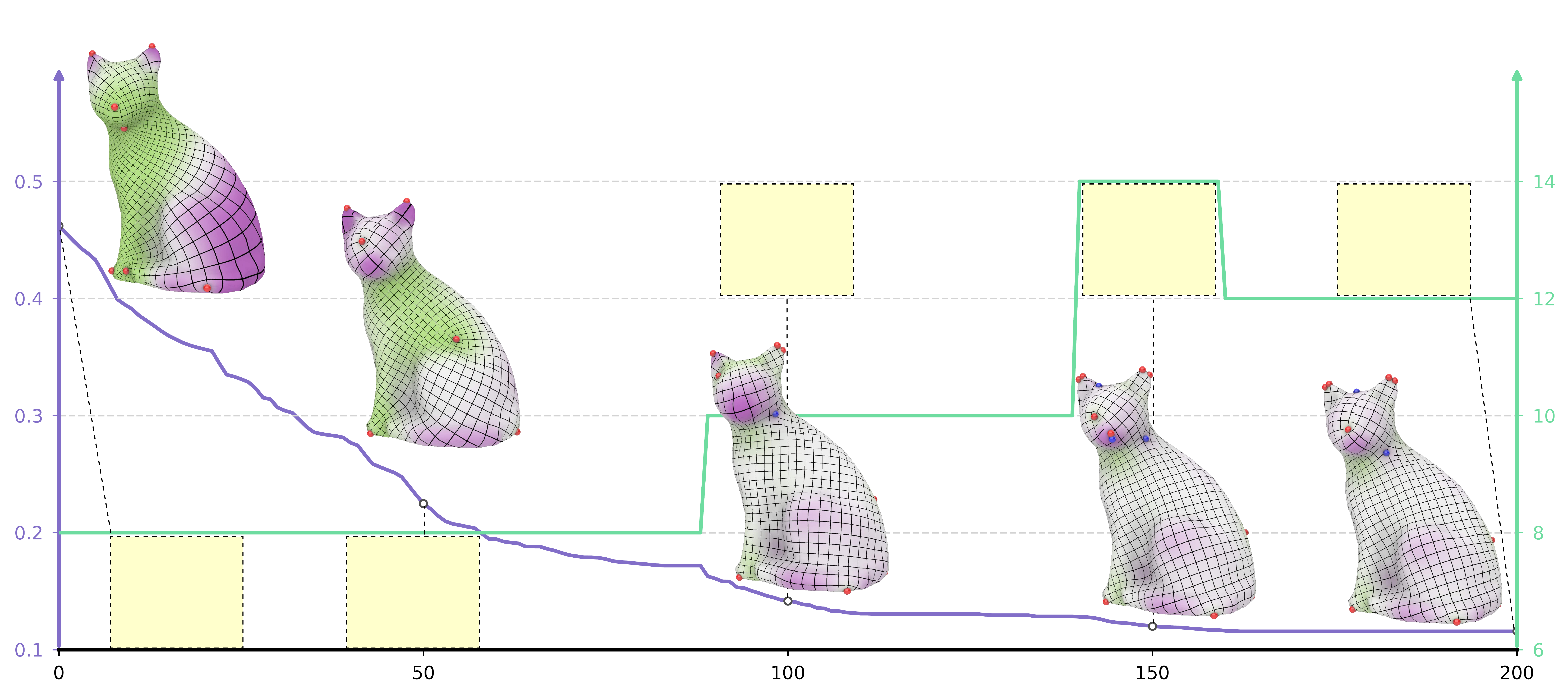}
    {
 \put(7.2,8.7){\small \#Iter: 0}
 \put(7.2,7.3){\small \(N_c=8\)}
 \put(7.2,5.9){\small \(N_0=0\)}
 \put(7.2,4.5){\small \(\mathcal{E}=0.462\)}
 \put(7.2,3.1){\small \(t=0.015s\)}
 
 \put(22.3,8.7){\small \#Iter: 50}
 \put(22.3,7.3){\small \(N_c=8\)}
 \put(22.3,5.9){\small \(N_0=0\)}
 \put(22.3,4.5){\small \(\mathcal{E}=0.225\)}
 \put(22.3,3.1){\small \(t=0.023s\)}
 
 \put(46.1,31.1){\small \#Iter: 100}
 \put(46.1,29.7){\small \(N_c=10\)}
 \put(46.1,28.3){\small \(N_0=0\)}
 \put(46.1,26.9){\small \(\mathcal{E}=0.142\)}
 \put(46.1,25.5){\small \(t=0.048s\)}
 
 \put(69.3,31.1){\small \#Iter: 150}
 \put(69.3,29.7){\small \(N_c=14\)}
 \put(69.3,28.3){\small \(N_0=5\)}
 \put(69.3,26.9){\small \(\mathcal{E}=0.120\)}
 \put(69.3,25.5){\small \(t=0.271s\)}
 
 \put(85.5,31.1){\small \#Iter: 200}
 \put(85.5,29.7){\small \(N_c=12\)}
 \put(85.5,28.3){\small \(N_0=0\)}
 \put(85.5,26.9){\small \(\mathcal{E}=0.116\)}
 \put(85.5,25.5){\small \(t=0.275s\)}

 \put(4,39){\textcolor{color1}{\small \(\mathcal{E}\)}}
 \put(97,39){\textcolor{color2}{\small \(N_c\)}}
 
    }
  \end{overpic}
  \vspace{-2mm}
  \caption{
  The graph plots the distortion $\mathcal{E}$ and the number $N_c$ of non-zero-angle cones vs. the number (\#Iter) of distortion change. Specifically, the iteration process of relocating cones is completely shown.
  We also report the number of zero-angle cones (denoted as \(N_0\)) and the runtime \(t\). 
  Five progressive results are shown, where the leftmost is the initialization and the rightmost is the result. 
  }
  \label{fig:pipeline}
\end{figure*}

\paragraph{Methodology}
Determining the cone configuration involves computing their discrete positions, angles, and numbers, which is a challenging discrete optimization problem.
We optimize each type of variable alternately until the distortion is less than the threshold $\epsilon_\text{tar}$ or the iteration number reaches the maximum (1000 in the experiments) (Alg.~\ref{alg:pipeline}).
Specifically, given a set of cones, we optimize their angles without changing positions (Sec.~\ref{subsubsec:solving_angles}) and reposition cones with fixed angles (Sec.~\ref{subsubsec:solving_position}).
Neither step increases distortion.
When freezing the angles and positions, we adaptively add (Sec.~\ref{subsubsec:adding_cone}) and remove (Sec.~\ref{subsubsec:removing_cone}) cones to change the number. 
The resulting cones induce a global seamless conformal parameterization.
Alg.~\ref{alg:pipeline} shows the pseudocode, and Fig.~\ref{fig:pipeline} shows an example. 



\IncMargin{0.5em}
\begin{algorithm}[t]
	\caption{Cone generation algorithm for genus-zero surfaces without boundaries}\label{alg:pipeline}
	\newcommand\mycommfont[1]{\footnotesize\ttfamily\textcolor{blue}{#1}}
	\SetCommentSty{mycommfont}
	\SetKwInOut{AlgoInput}{Input}
	\SetKwInOut{AlgoOutput}{Output}

    \SetKwFunction{GetNonSpStandardPatch}{GetNonSpStandardPatches}
    \SetKwFunction{Init}{Initializations}
    \SetKwFunction{SolveAnglesWithFixedPosition}{SolveAnglesWithFixedPositions}
    \SetKwFunction{OPFA}{UpdatePositionsWithFixedAngles}
    \SetKwFunction{AddingCones}{AddingCones}
    \SetKwFunction{RemovingCones}{RemovingCones}
    \SetKwRepeat{Do}{do}{while}
	\AlgoInput{A genus-zero surface $\mM$ without boundaries, a target distortion \(\epsilon_\text{tar}\), and the maximum iteration number \(N_\text{iter}\).}
	\AlgoOutput{A small set of integer-constrained cones which induce a global seamless conformal parameterization.}
    $\Init();$ \, \tcc{Sec.~\ref{subsubsec:adding_cone}} \
    \(i \gets 1\)\;
    \While{\(i<N_\text{iter}\) and \(\mathcal{E} > \epsilon_\text{tar}\) }
    {
        $\,\AddingCones();$ \, \tcc{Sec.~\ref{subsubsec:adding_cone}} \
            \SolveAnglesWithFixedPosition(); \, \tcc{Sec.~\ref{subsubsec:solving_angles}} \
            \OPFA(); \, \tcc{Sec.~\ref{subsubsec:solving_position}} \
        \RemovingCones(); \, \tcc{Sec.~\ref{subsubsec:removing_cone}}\
        \(i \gets i+1\)\;
    }
\end{algorithm}
\DecMargin{0.5em}



\subsection{Efficient optimization with fixed numbers}\label{sec:moving}
\paragraph{Formulation}
Given a set of cones, we perform the following constrained optimization to update angles and positions without changing the number:
\begin{equation}
\begin{aligned}\label{original_problem}
    \min_{\mathbf{k}^\text{tar},\mathbf{u},\mathbf{z}} \quad & \textblue{\mathcal{E}^2}\\ 
    s.t. \quad &\sum_{i=1}^{N} k_i^\text{tar}=4\pi,\\
    &\mathbf{k}^\text{tar} = \frac{\pi}{2}\mathbf{z}, \mathbf{z} \in \mathbb{Z}^\text{N},\\
    &\mathbf{L}\mathbf{u}=\mathbf{k}^\text{tar}-\mathbf{k}^\text{ori},
\end{aligned}
\end{equation}
where $\mz$ is an $N$-dimensional vector whose each entry is an integer.

\subsubsection{Solving angles with fixed positions}\label{subsubsec:solving_angles}
\paragraph{Key idea for high efficiency}
If we directly solve the problem~\eqref{original_problem}, the number of integer variables is the same as the number $N$ of vertices, which is time-consuming.
In practice, reducing the number of integer variables is a promising way to improve efficiency.
We observe that \(\mathbf{k}^\text{tar}\) is a sparse vector with known positions of non-zero entries under fixed positions of cones. Specifically, the number of non-zero entries is the same as that of cones.
Thus, we can leverage the Yamabe equation to express \(\mathbf{u}\) in terms of those non-zeros of \(\mathbf{k}^\text{tar}\), thereby allowing us to reduce~\eqref{original_problem} to an equivalent problem, whose number of integer variables is similar to the cone number.

\paragraph{Solution of the Yamabe equation}
Given fixed positions of cones, we sequentially collect the non-zero entries of $\mz$ into a vector \(\mz^\text{int}=(z_{c_1},...,z_{c_{N_c}})^T\).
We can use a linear transformation to connect $\mz^\text{int}$ and $\mz$: $\mz=\mathbf{T}\mz^\text{int}$, where \(\mathbf{T}\) is an \(N \times N_c\)-dimensional matrix defined by \(\mz^\text{int}\). 
Accordingly, the Yamabe equation becomes:
\begin{equation}\label{yamabe_with_integers}
    \mathbf{L}\mathbf{u}=\frac{\pi}{2}\mathbf{z}-\mathbf{k}^\text{ori} = \frac{\pi}{2}\mathbf{T}\mz^\text{int}-\mathbf{k}^\text{ori}.
\end{equation}

\tb{The solution to the Yamabe equation in \eqref{yamabe_with_integers} is unique up to a constant global scaling factor. Formally, since} the rank of \(\mathbf{L}\) is \((N-1)\) and its null space is spanned by an \(N\)-dimensional all-ones vector \(\mathbf{1}_{N}\), the general solution of~\eqref{yamabe_with_integers} is
\begin{equation}\label{original_general}
    \mathbf{u}=\mathbf{u}^\text{par}+a\mathbf{1}_{N},
\end{equation}
where \(\mathbf{u}^\text{par}\) is a particular solution and \(a\) is a free continuous variable.
Without loss of generality, we select a mesh vertex \(v_p\) that is not a cone singularity and set its log conformal factor \(u_p\) to zero to find a particular solution. 
This gives:
\begin{equation}\label{u_with_k}
\widehat{\mathbf{L}} \mathbf{u}^\text{par} = \mathbf{M}(\frac{\pi}{2}\mathbf{T}\mathbf{z}^\text{int}-\mathbf{k}^\text{ori}),
\end{equation}
where the non-singular matrix \(\widehat{\mathbf{L}} \) differs from \(\mathbf{L}\) only in \(p\)-th row and \(p\)-th column, with the diagonal element set to one while other entries are zeros and \(\mathbf{M}\) is an identity matrix except that its \(p\)-th diagonal entry is zero. 
Then, by combining~\eqref{original_general} and~\eqref{u_with_k}, we have
\begin{equation}\label{reduce_equation}
\begin{aligned}
    \mathbf{u}&=\mathbf{u}^\text{par}+a\mathbf{1}_N\\
    &=\widehat{\mathbf{L}}^{-1}\mathbf{M}(\frac{\pi}{2}\mathbf{T}\mathbf{z}^\text{int}-\mathbf{k}^\text{ori})+a\mathbf{1}_N.
\end{aligned}
\end{equation}

\paragraph{Optimization}
After substituting~\eqref{reduce_equation} into the problem~\eqref{original_problem}, optimizing cone angles under fixed positions can be reduced to
\begin{equation}\label{small_scale_problem}
\begin{aligned}
    \min_{\mathbf{z}^\text{int},a}\ & \Vert\mathbf{A}^{\frac{1}{2}}(\frac{\pi}{2}\widehat{\mathbf{L}}^{-1} \mathbf{M} \mathbf{T} \mathbf{z}^\text{int}-\widehat{\mathbf{L}}^{-1} \mathbf{M}\mathbf{k}^\text{ori} + a\mathbf{1}_N)\Vert_2^2\\
    s.t. \ & \sum_{i=1}^{N_c} z_{c_i}=8,\\
    &\mathbf{z}^\text{int} \in \mathbb{Z}^{N_c}.
\end{aligned}
\end{equation}
This problem has $N_c$ integer variables and one continuous variable.
%
To perform further simplification, we can express \(z_{c_{N_c}}\) as a linear combination of \(z_{c_1}, \ldots, z_{c_{N_c-1}}\) using the Gauss-Bonnet theorem, i.e., $\sum_{i=1}^{N_c} z_{c_i}=8$.
Then, this linear topology constraint can be eliminated, and we convert the problem~\eqref{small_scale_problem} to one
\textblue{without linear constraints}.
This allows us to apply existing mixed-integer solvers to optimize the objective and determine the cone singularity angles. 

\paragraph{Active pruning}
As the number of cones is too large, solving~\eqref{small_scale_problem} using an existing mixed-integer solver becomes inefficient. Therefore, if the number of cones is greater than $N_g$ (30 in our experiment), we apply active pruning by collecting \(N_g\) cones to solve angles, while the angles of all other cones are kept fixed.
Since we only solve angles after each adding cone step, it is necessary to always select the newly added cones. 
\textblue{As newly added cones primarily influence nearby cones, we select the cones closest to them to collect a total of \(N_g\) cones.}


\subsubsection{Optimizing positions with fixed angles}\label{subsubsec:solving_position}
Given the number and angles of cones, we adopt a similar cone relocation strategy to \cite{li2023efficient}, where cones are moved across the mesh to reduce area distortion progressively.
\textblue{In our setting, the cone position is constrained to mesh vertices.}
\paragraph{A revisit of \cite{li2023efficient}}
According to \cite{li2023efficient,sharp2018variational}, to reduce area distortion under the Yamabe equation, a Lagrangian function is constructed to derive the gradient.
The normal directional derivative \(\nabla_n \mathcal{E}\) of the area distortion with respect to the moving direction \(n\) of a varying-angle cone $c$ is:
\begin{equation}\label{E_gradient}
    \nabla_n \mathcal{E}=\frac{1}{2\mathcal{E}}(u^2-\frac{\partial u}{\partial n}\frac{\partial h}{\partial n}),
\end{equation}
where \(h\) serves as the Lagrangian multiplier and is the solution of
\begin{equation}\label{lm_h}
\begin{aligned}
    \Delta h&=-2u,\ \text{on}\ \mathcal{M}\backslash\bigcup_{c\in\mathcal{C}} \partial\delta_c,\\
    h&=0,\ \text{on}\ \bigcup_{c\in\mathcal{C}} \partial\delta_c.
\end{aligned}
\end{equation}
Here, \(\partial \delta_c\) is the boundary of a small neighborhood of \(c\). 
\( h, \nabla_n \mathcal{E} \) can be obtained by discretizing and solving~\eqref{lm_h}~\cite{li2023efficient}.
Therefore, the gradient \(\text{grad}_c\mathcal{E}\) for each cone \(c\) can be derived using \(\nabla_n\mathcal{E}\) (see Equation~(11) in~\cite{li2023efficient}).
\textblue{The cone is then moved to its adjacent vertex in the direction of \(-\text{grad}_c\mathcal{E}\) with the greatest magnitude, until \(\mathcal{E}\) no longer decreases.}

\paragraph{Our case}
In~\cite{li2023efficient}, the cone angles are continuous variables that are continuously updated during movement.
However, our angles remain unchanged during the movement, leading to a different Lagrangian function and \(h\) is the solution of
\begin{align}\label{my_h}
    \Delta h&=-2u,\ \text{on}\ \mathcal{M}.
\end{align}
We add the detailed discussion in 
\textblue{supplements.}
As shown in Fig.~\ref{fig:different_h}, if we still use~\eqref{lm_h} to derive the movement direction, some cones may get stuck at the initial positions, 
indicating that the solution \(h\) of~\eqref{lm_h} is no longer practical to our problem. 
Instead, using~\eqref{my_h} to compute the directional derivative leads to effective movement.

\begin{figure}[t]
  \centering
  \begin{overpic}[width=0.99\linewidth]{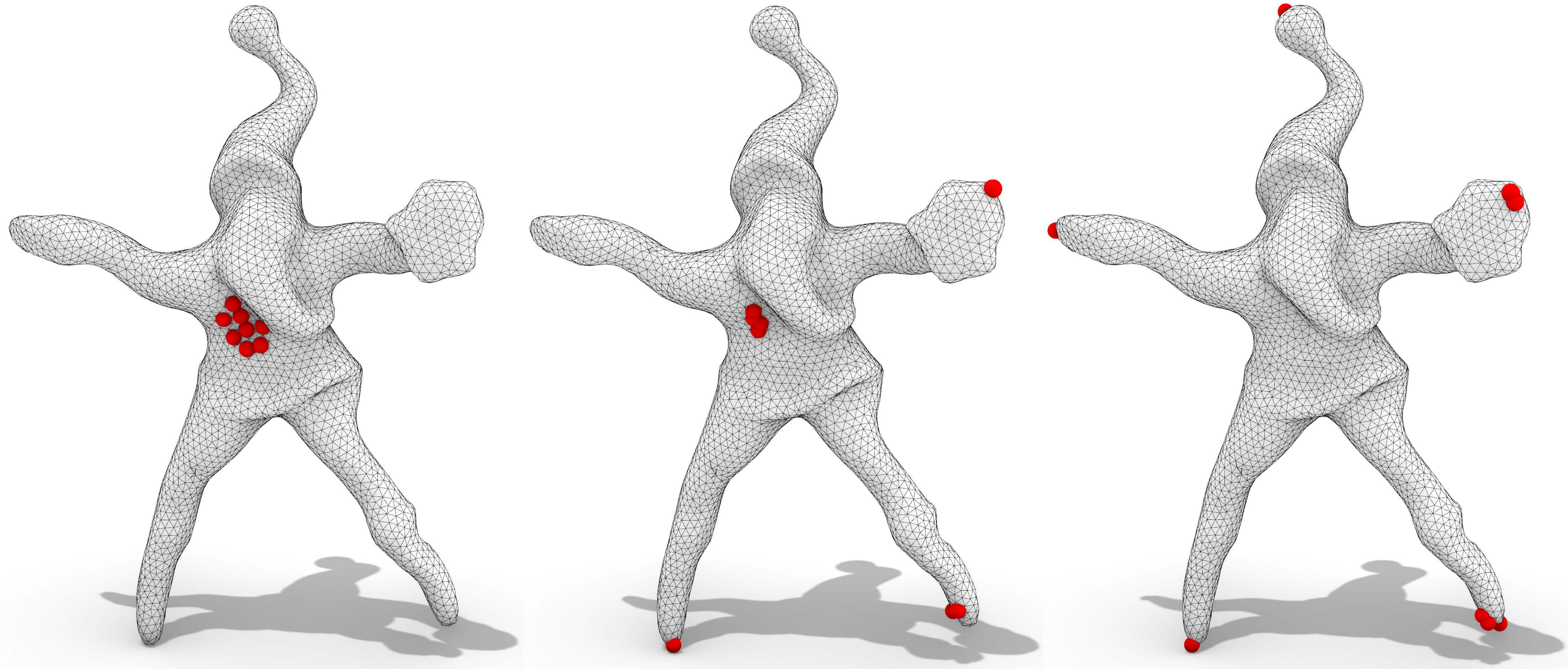}
    {
 \put(8,-3.5){\small Initialization}
 \put(45,-3.5){\small Using~\eqref{lm_h}}
 \put(78,-3.5){\small Using~\eqref{my_h}}
    }
  \end{overpic}
  \vspace{-1mm}
  \caption{
  Different methods for computing \(h\).
  After placing eight cones with angles of \(\frac{\pi}{2}\) on the belly of the Santa model (left), cones are moved along the negative gradient directions computed using \(h\) from \eqref{lm_h} and \eqref{my_h}. 
  }
  \label{fig:different_h}
\end{figure}

\paragraph{Moving cones}
We constrain each cone to only move toward its neighboring vertices along the computed direction, similar to~\cite{li2023efficient}. 
To sufficiently reduce the distortion while ensuring high efficiency, we perform movement using the following steps in one iteration: (1) update all cones simultaneously; (2) if Step (1) does not decrease distortion, update the cones one by one.
We iteratively conduct the movement until there is no distortion reduction.

\subsubsection{Algorithm analysis}
\paragraph{Non-increasing distortion} 
In these two steps, our algorithm does not increase distortion. 
After solving angles with fixed positions, the optimized solution will not lead to a higher distortion than the previous. 
If the distortion increases within repositioning, we reject the update and retain the previous solution.
Hence, the distortion does not increase (Fig.~\ref{fig:pipeline}).

\paragraph{Performance}
When solving angles with~\eqref{small_scale_problem}, we need to evaluate \(\frac{\pi}{2}\widehat{\mathbf{L}}^{-1} \mathbf{M} \mathbf{T}\) and \(\widehat{\mathbf{L}}^{-1} \mathbf{M}\mathbf{k}^\text{ori}\). 
To enable pre-factorization of \(\widehat{\mathbf{L}}\), we fix the vertex \(v_p\) on the mesh.
As a result, \(\widehat{\mathbf{L}}\) remains fixed and can be pre-factorized. Additionally, since \(\mathbf{T}\) has a small number of columns and the system is solved infrequently, solving angles is computationally efficient.  
During repositioning, the main cost comes from evaluating \(h\). Unlike the discrete formulation of~\eqref{lm_h}, our formulation of~\eqref{my_h} has a fixed coefficient matrix, which allows for one-time factorization without the need for Cholesky updates in subsequent iterations~\cite{li2023efficient,myles2012global}.  
Thus, our method is overall highly efficient.

\begin{figure}[t]
  \centering
  \begin{overpic}[width=0.99\linewidth]{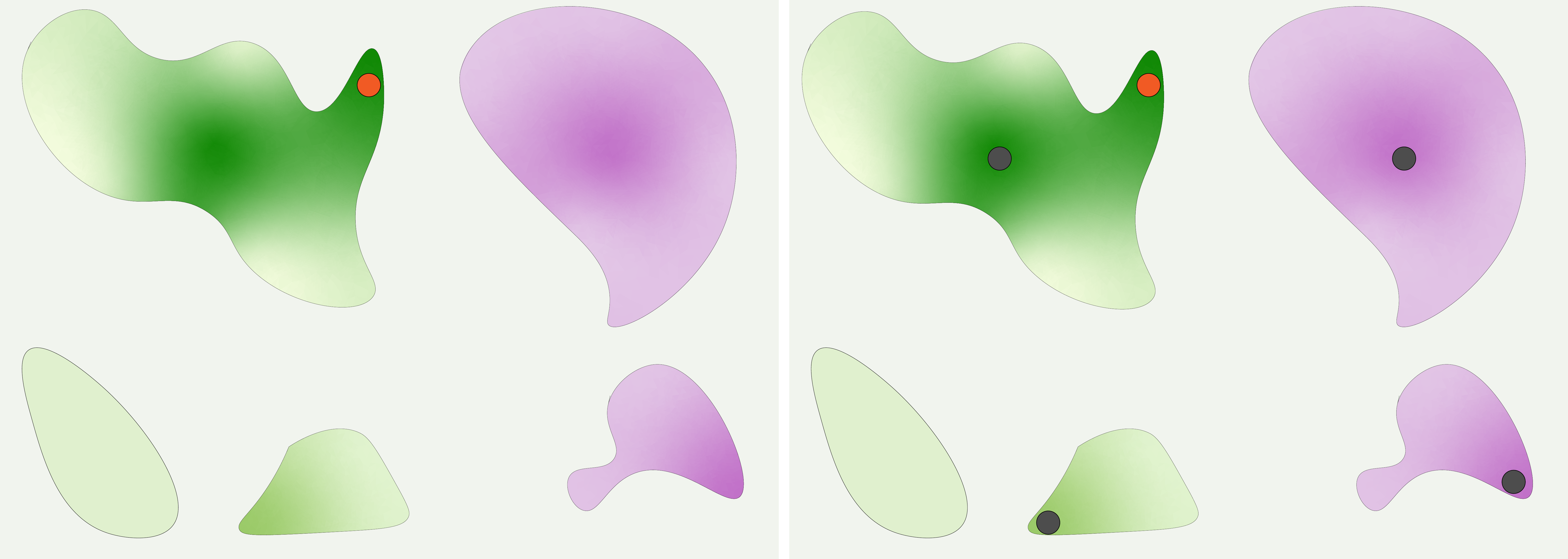}
    {
  \put(17,-3.5){\small Before adding}
  \put(67,-3.5){\small After adding}
    }
  \end{overpic}
  \vspace{-1mm}
  \caption{
  Adding cones.
The green (\(f > 0\)) and purple (\(f < 0\)) regions are the computed branches, with darker shades indicating larger \(|f|\). We place four cones (gray points) at the locations on the top branches, ranked by \(\mathcal{E}_\text{branch}\), where the scalar field \(|f|\) reaches its maximum, excluding positions that are already occupied by existing cones (orange points).
  }
  \label{fig:add-cone}
\end{figure}

\subsection{Adaptive Adjustment of cone numbers}\label{sec:number}
\subsubsection{Adding cones}\label{subsubsec:adding_cone}
It is possible to increase the number of cone singularities to lower distortion. The effectiveness of introducing additional cones via a branch-driven strategy has been confirmed by~\cite{li2023efficient,chai2018sphere}, and we adopt a similar approach.

\paragraph{A revisit of the branch-driven method}
Let \( f(v) \) be a scalar field defined on the mesh vertices, which serves as an indicator of parameterization distortion. %
We set a threshold \( f^\text{thres} > 0 \), and identify all vertices \( v \) such that \( |f(v)| > f^\text{thres} \). Starting from each such vertex, we perform a depth-first search to find several vertex sets \( \{S(v)\} \) satisfying the following conditions:
\begin{itemize}
    \item For each \( v \in S(v) \), we have \( |f(v)| > f^\text{thres} \);
    \item Each set \( S(v) \) is simply connected on the mesh;
    \item All values of \( f(v) \) within a set share the same sign.
\end{itemize}
This process yields a collection of simply connected vertex sets, called \emph{branches}. 
The distortion of each branch is measured by the area-weighted norm of the function~\( f \), given by:
\begin{equation}
\mathcal{E}_\text{branch}=\sum_{v \in S(v)} \text{Area}(v) \, |f(v)|^2.
\end{equation}
These branches are sorted in descending order of \(\mathcal{E}_\text{branch}\). To introduce \( N_a \) new cones, the vertex with the largest \( |f(v)| \) (that is not already a cone) is added from each of the top \( N_a \) branches, as shown in Fig.~\ref{fig:add-cone}.

\paragraph{Adding cones for initialization}
To determine the initial cone positions, we use Gaussian curvature as the scalar field \(f\) to apply the branch-driven strategy.
The number \( N_a \) of cones is set to eight.

\paragraph{Adding cones during iterations}
To determine new cones in each iteration, we follow~\cite{li2023efficient} by setting the scalar field $f$ as \(\mathbf{u}\).
Unlike~\cite{li2023efficient}, our cone angles are restricted rather than arbitrary, so adding a single cone only leads to a slight reduction in distortion. 
However, adding too many cones at once may lead to redundancy.
Thus, we insert more cones when the current distortion is large, and fewer when it is small.
Specifically, we insert \(\min\{m,10\}\) cones if the distortion exceeds \(m\epsilon_\text{tar}\) and the cone number is greater than \(N_g\).  
Otherwise, only one cone is added to avoid introducing unnecessary cones.

The newly inserted cones are initialized with zero angles before angle optimization. After optimization, some of them may still have zero angles, but we keep them as they may become effective in subsequent iterations.

\begin{figure}[t]
  \centering
  \begin{overpic}[width=0.7\linewidth]{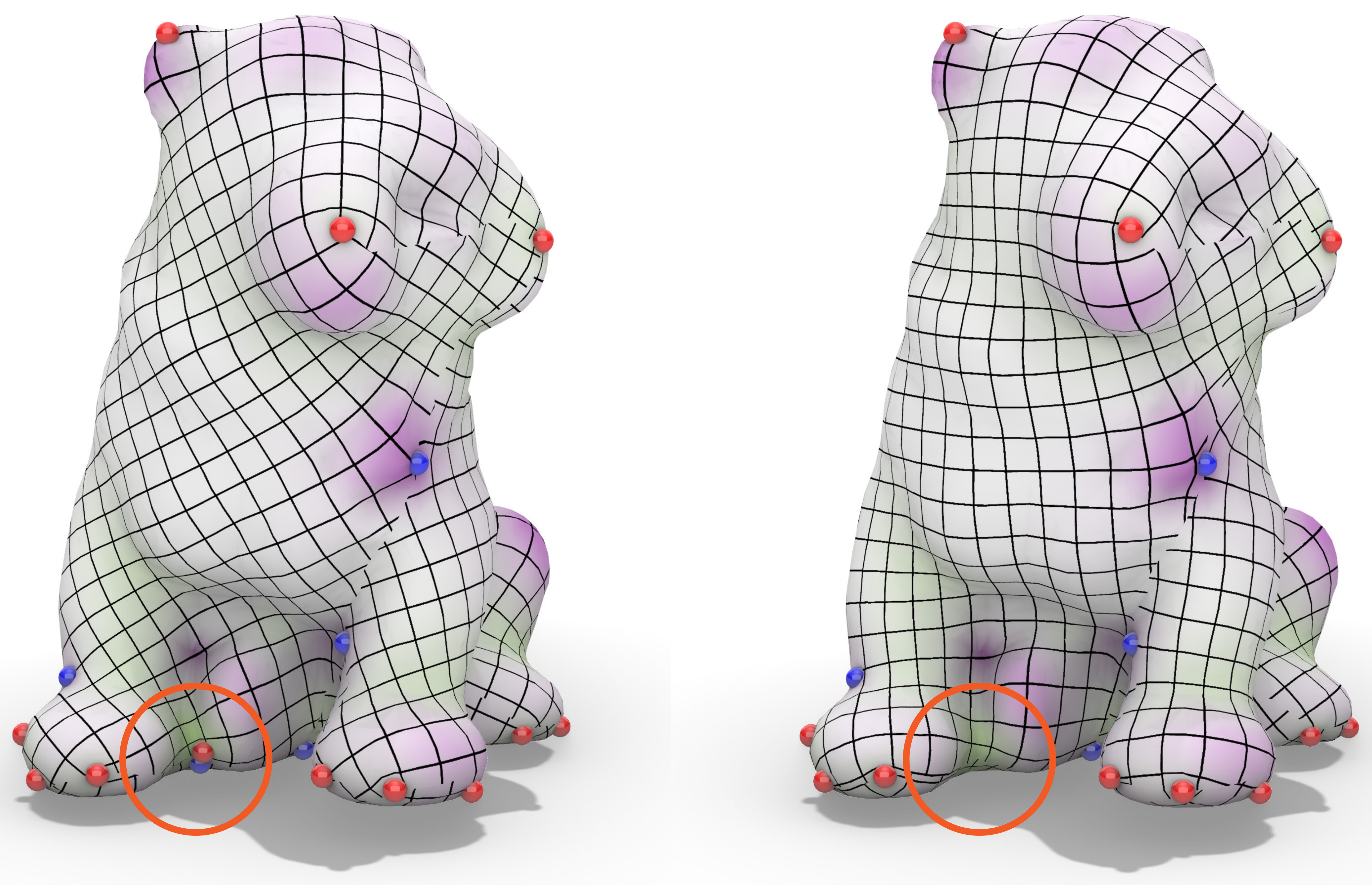}
    {
 \put(7,-4){\small Before removing}
 \put(64,-4){\small After removing}
    }
  \end{overpic}
  \vspace{2mm}
  \caption{
  Removing cones. After removing a pair of cones (in the orange circle), the distortion increases slightly from 0.1176 to 0.1181, while the number of cones decreases from 26 to 24.
  }
  \label{fig:remove-cone}
\end{figure}

\subsubsection{Removing cones}\label{subsubsec:removing_cone}
As illustrated in Fig.~\ref{fig:remove-cone}, parts of cone pairs with opposite angles tend to be ineffective in reducing distortion when they are located too close to each other. Therefore, during the iteration, we detect such pairs and attempt to remove them, similar to \cite{zhang2023practical}. 
Specifically, we collect cone pairs with opposite angles whose distance on the mesh (measured as the number of edges between them) is less than \(5\times10^{-4}N\), as shown in Fig.~\ref{fig:remove-cone}.
If removing such a pair results in a relative distortion increase smaller than a threshold \( \eta\), we remove the pair; otherwise, we retain it.
Since removing cones may lead to an increase in energy, to guarantee the convergence of our algorithm, we dynamically update \(\eta\) during the optimization.
Each time a pair of cones is successfully removed, we update \( \eta\) to \( 0.9\eta \). In the limiting case, \(\eta\) approaches zero, at which point we stop removing cones. Hence, the convergence of our algorithm is still guaranteed.
In the experiments, the initial \(\eta\) is 10\%.

\subsubsection{Gauss-Bonnet theorem}
\textblue{When adding cones, the newly inserted cones are initialized with zero angles. When removing cones, we always remove pairs with opposite angles. Therefore, the total cone angle remains unchanged during both insertion and removal, ensuring the Gauss–Bonnet constraint is preserved throughout.}

\section{Methods for nonzero genus surfaces} 
\subsection{Seamless conformal parameterization}
For a closed mesh with genus \(g > 0\), a global seamless conformal parameterization must satisfy not only the previously mentioned constraints, but also an additional set of holonomy constraints: the holonomy angles for \(2g\) non-contractible handles and tunnels must be integer multiples of \(\frac{\pi}{2}\)~\cite{myles2012global}. In other words, the integral of geodesic curvature along each such loop must be a multiple of \(\frac{\pi}{2}\):
\begin{equation}\label{continuous_holonomy_constraint}
    \int_{\gamma_i}k^g(s)ds=\frac{\pi}{2}r_i,\ r_i\in\mathbb{Z},\ i=1,\ldots,2g,
\end{equation}
where \(\gamma_i\) is the \(i\)-th non-contractible homology loop and \(k^g\) is the geodesic curvature of the original mesh.

In the discrete case, it suffices to require that the integral of geodesic curvature on left side of the cut is a multiple of \(\frac{\pi}{2}\)~\cite{Li2022int}.
So if considering the \(i\)-th loop on the mesh, which contains vertices \(G_i\) on the left side of the cut, the discrete form of the constraint (\ref{continuous_holonomy_constraint}) is
\begin{equation}\label{discrete_holonomy_constraint}
    \sum_{j\in G_i}\mathbf{L}_j^\text{l}\mathbf{u}=\frac{\pi}{2}r_i-\sum_{j\in G_i}k^g_{i,j},
\end{equation}
where \(\mathbf{L}_j^\text{l}\) denotes the Laplacian matrix computed using only the vertex \(v_j\) and its 1-ring vertices on the left side of the loop and \(k^g_{i,j}\) is the geodesic curvature of the vertex in \(G_i\).


The resulting system by combining (\ref{discrete_holonomy_constraint}) with (\ref{original_problem}) becomes over-constrained and has no solution, indicating that a global seamless conformal parameterization does not exist in general. 
Thus, we instead seek a rotationally seamless parameterization, in which the log conformal factors on the two sides of each loop are allowed to differ, same goal with~\cite{zhang2023practical,Li2022int}. 
To achieve this, we cut the mesh along the loops corresponding to the handles and tunnels, and assign independent variables to the duplicated vertices.

\subsection{Pipeline}
\begin{figure}[t]
  \centering
  \begin{overpic}[width=0.75\linewidth]{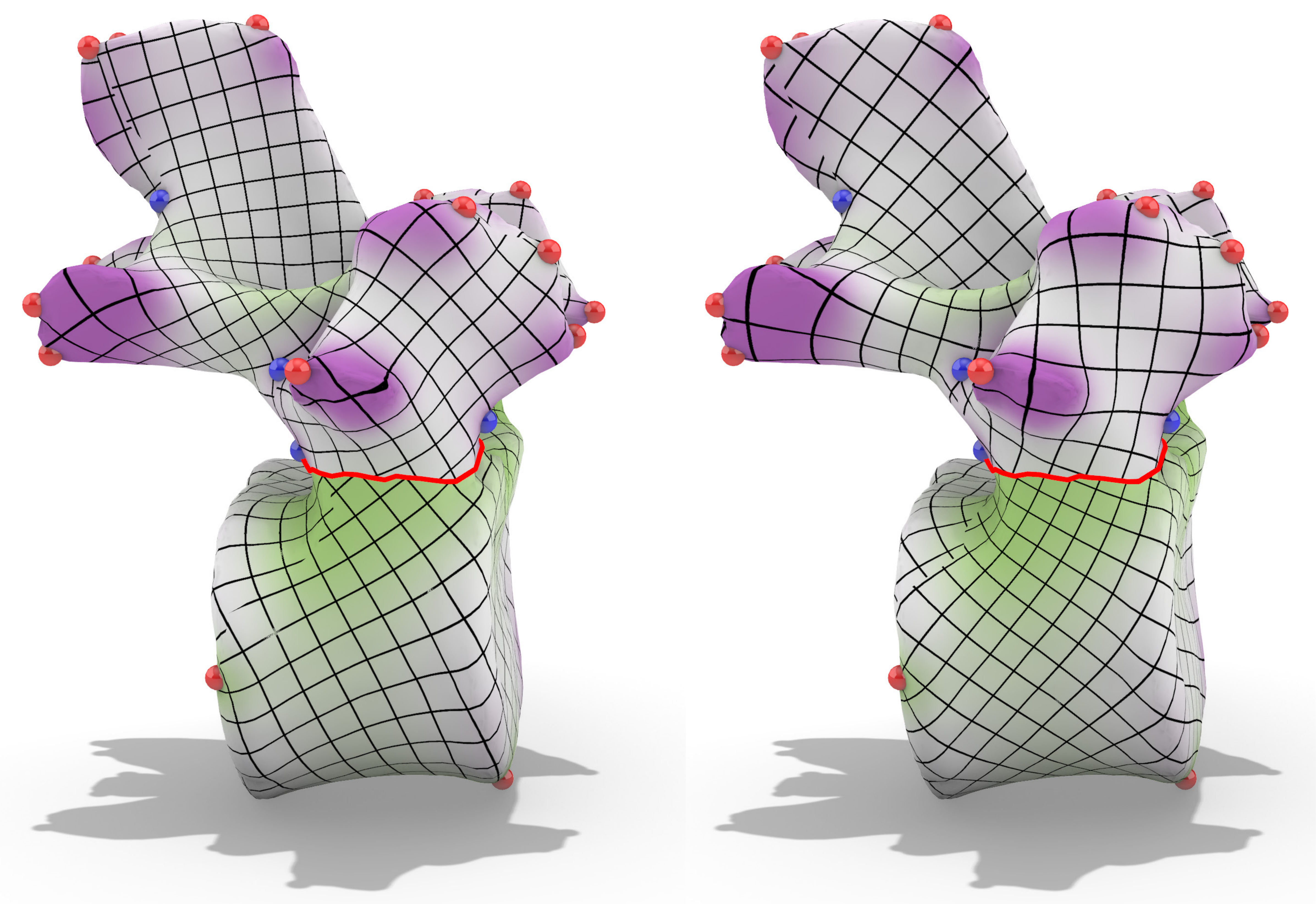}
    {
 \put(14,-3.5){\small W/O constraint}
 \put(66,-3.5){\small With constraint}
    }
  \end{overpic}
  \vspace{2mm}
  \caption{
  (Left) Without holonomy angle constraints, the texture across the cut (red) will mismatch. (Right) In contrast, with these constraints, the result is rotationally seamless.
  }
  \label{fig:holonomy}
\end{figure}

Our pipeline contains two steps, similar to~\cite{zhang2023practical}.
First, we ignore the holonomy angle constraints and solve~\eqref{small_scale_problem} to determine the positions and angles of the cone singularities.
Different from Sec.~\ref{sec:method}, the topology constraint becomes \(\sum_{i=1}^{N_c}z_{c_i}=8(1-g)\) and the number of initial cones needs to be adjusted to \(|8(1-g)|\) accordingly. 
In the second step, we cut the mesh along the non-contractible homology loops and solve for a new log conformal factor while keeping the cone configuration fixed, such that the additional constraints are satisfied and the difference of log conformal factors on either side of the cuts remains small (Fig.~\ref{fig:holonomy}).
Next, we will focus on discussing the second step.

\subsection{Solving the second step}
The enumeration-driven rounding strategy~\cite{zhang2023practical} for computing the integer vector \(\mathbf{r}=(r_1,...,r_{2g})^T\) is computationally expensive.  
Instead, we continue to use our variable reduction approach. 

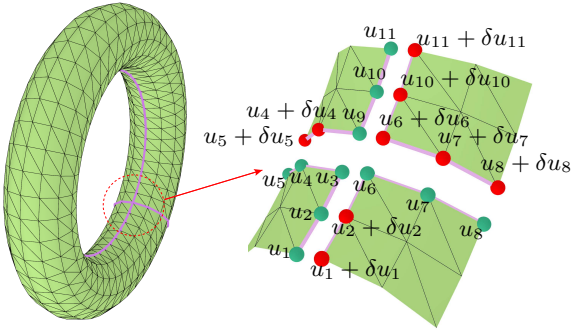
\begin{figure}[t]
  \centering
  \begin{overpic}[width=0.8\linewidth]{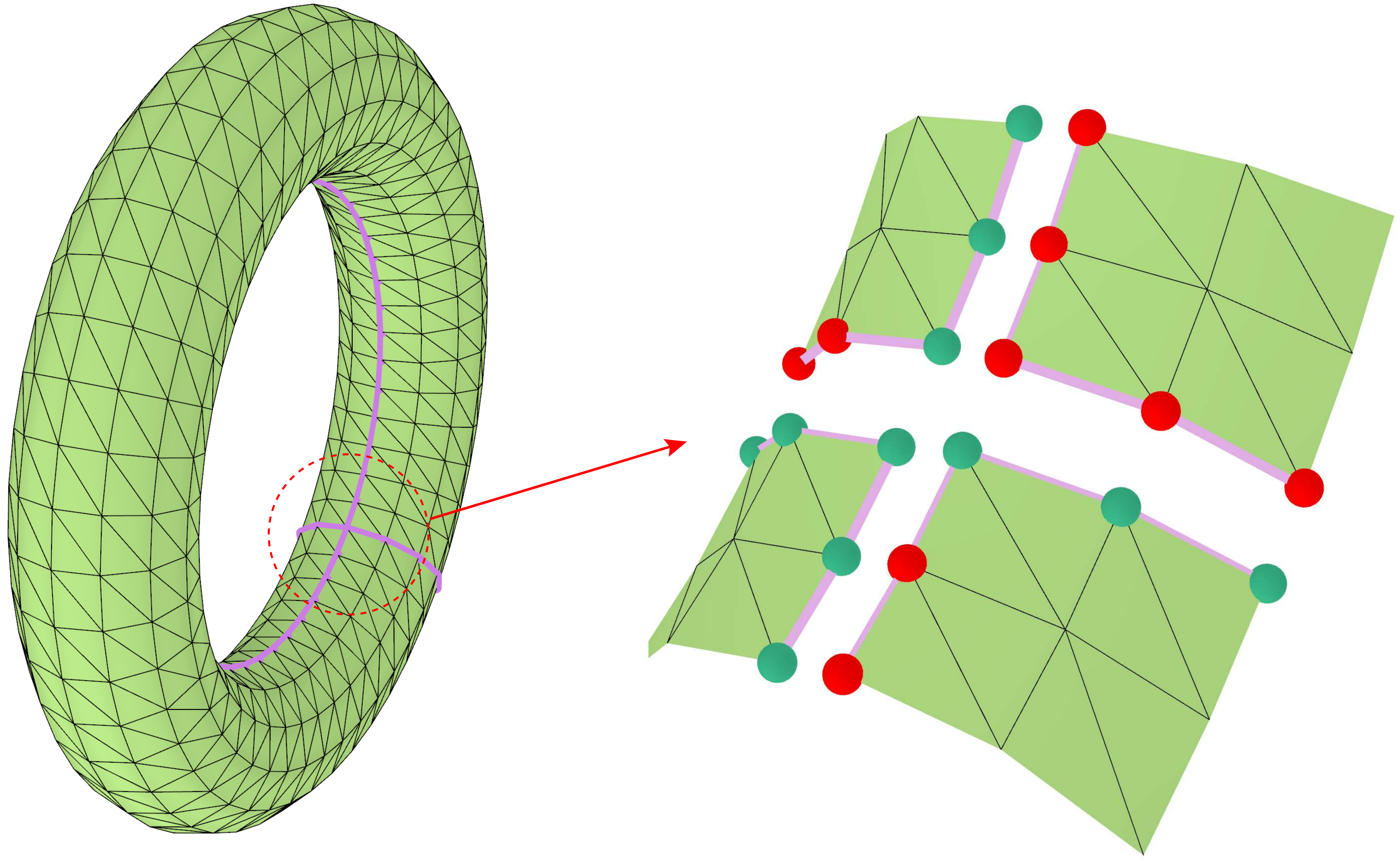}
    {
 \put(50,14){\small \(u_1\)}
 \put(54,21){\small \(u_2\)}
 \put(59,27.5){\small \(u_3\)}
 \put(54,27.5){\small \(u_4\)}
 \put(49,27){\small \(u_5\)}
 
 \put(58,10){\small \(u_1+\delta u_1\)}
 \put(62,18){\small \(u_2+\delta u_2\)}
 \put(66,26){\small \(u_6\)}
 \put(76,22.5){\small \(u_7\)}
 \put(86,17.5){\small \(u_8\)}

 \put(71,38.5){\small \(u_6+\delta u_6\)}
 \put(82,35){\small \(u_7+\delta u_7\)}
 \put(90,30){\small \(u_8+\delta u_8\)}
 \put(75,46){\small \(u_{10}+\delta u_{10}\)}
 \put(78,53.5){\small \(u_{11}+\delta u_{11}\)}
 
 \put(64,39.5){\small \(u_9\)}
 \put(66,47.5){\small \(u_{10}\)}
 \put(68,55){\small \(u_{11}\)}
 \put(46,40){\small \(u_4+\delta u_4\)}
 \put(38,35){\small \(u_5+\delta u_5\)}
    }
  \end{overpic}
  \vspace{-1mm}
  \caption{
  New variable assignment.
  For the red vertices on the cut boundaries, we introduce a nonzero \(\delta_u\), while for the other green vertices, \(\delta_u=0\).
  }
  \label{fig:assemble_delta_u}
\end{figure}

\paragraph{Null space of constraint matrix}
To enable this, we assign each vertex with a continuous variable \(u\), assemble all the constraints \eqref{yamabe_with_integers} and \eqref{discrete_holonomy_constraint} into a single linear system 
\begin{equation}\label{original_constraint_system_for_highgenus}
\mathbf{L}_\text{h}\mathbf{u}_\text{h}=\mathbf{b}_\text{h}
\end{equation}
and compute the null space \(\text{Null}(\mathbf{L}_\text{h})\) of the matrix \(\mathbf{L}_\text{h}\).
This step is indispensable, as the ability to express \(\mathbf{u}\) in terms of the integer variables \(\mathbf{r}=(r_1,\ldots,r_{2g})\) critically depends on the structure of \(\text{Null}(\mathbf{L}_\text{h})\).

\paragraph{Dimension of null space}
To analyze the null space, we first examine its dimension.
After cutting the mesh, each loop leads to two duplicated vertices on either side of the cut loops, except at the intersection points between handles and tunnels, where four vertices are produced. 
Denote that the number of vertices on the \(i\)-th handle and tunnel loops are \(G_i^h\) and \(G_i^t\), respectively. Then, the number of newly introduced variables after cutting is
\begin{equation}
\begin{aligned}
&\sum_{i=1}^g \left(2(G_i^h - 1 +G_i^t - 1) + 4  - (G_i^h + G_i^t- 1) \right)\\
&= \sum_{i=1}^g\left(G_i^h+G_i^t+1\right).
\end{aligned}
\end{equation}
However, only \(2g\) additional constraints are introduced, implying that the dimension of \(\text{Null}(\mathbf{L}_\text{h})\) is \(1 + \sum_{i=1}^g\left(G_i^h+G_i^t-1\right)\).

\paragraph{Numerical methods}
Numerical solutions for computing the null space of a matrix generally rely on matrix decomposition, which requires substantial computational resources and may suffer from numerical issues~\cite{coleman1986null,coleman1987null,gotsman2008computation}.
Therefore, we propose a practical method to explicitly construct the null space, which is more efficient and robust than numerical methods.



\paragraph{Explicit construction}\label{para:explicit_construction}
\textblue{Following the approach used for the genus-zero case, the Yamabe equation imposes $N-1$ constraints, and we introduce $N$ continuous variables. As a result, the constraint matrix is rank-deficient by one. This observation inspires us to treat the higher-genus case in a similar manner: due to the presence of $2g$ additional constraints, we likewise aim to introduce $2g$ additional variables so that the constraint matrix remains rank-deficient by one, ensuring its null space is trivial. To achieve this, we carefully design the variable assignment along the cuts, while preserving the assignments of all interior vertices.}
\textblue{We set many split vertices along the cuts to share the same variable, and additionally introduce an offset between them, similar to~\cite{tong2006designing}.}

As illustrated in Fig.~\ref{fig:assemble_delta_u}, we assign the same variable to two vertices on either sides, except assigning three variables to four corners at the intersection by letting one pair of adjacent corners share the same variable.
\tb{In other words, the conformal factors on both sides of each cut are represented by the same variable \(u_i\), 
except at the intersection points, where three variables (e.g., \(u_3, u_6, u_9\) in Fig.~\ref{fig:assemble_delta_u}) are used. 
Compared with the original uncut mesh, only two additional variables are introduced at each intersection point. 
Since there are \(g\) such intersections in total, the system therefore introduces \(2g\) additional variables. 
However, since the conformal factors on the two sides of each cut are inherently different, 
we introduce a sparse variable vector \(\delta \mathbf{u}\) as a modification to \(\mathbf{u}\). 
The entries of \(\delta \mathbf{u}\) are set to zero at all interior vertices, 
on the left-side boundary vertices, and at the three corner vertices along the boundary.}
Hence, the constraints become:
\begin{enumerate}
    \item For constraints on interior vertices \(v_i\): \(\mathbf{L}_i(\mathbf{u}+\mathbf{\delta u})=\frac{\pi}{2}z_i-k_i^\text{ori}\),
    \item For constraints on boundary vertices \(v_i^0,v_i^1,v_i^2,v_i^3\) of the intersection: \((\mathbf{L}_i^0+\mathbf{L}_i^1+\mathbf{L}_i^2+\mathbf{L}_i^3)(\mathbf{u}+\mathbf{\delta u})=\frac{\pi}{2}z_i-k_i^\text{ori}\), where \(\mathbf{L}_i^j\) is the Laplacian matrix only considering the vertex \(v_i^j\) and its neighbours.
    \item For constraints on boundary vertices \(v_i^\text{l},v_i^\text{r}\) on opposite side: \((\mathbf{L}_i^\text{l}+\mathbf{L}_i^\text{r})(\mathbf{u}+\mathbf{\delta u})=\frac{\pi}{2}z_i-k_i^\text{ori}\), where \(\mathbf{L}_i^\text{l},\mathbf{L}_i^\text{r}\) is the Laplacian matrix only considering the vertex \(v_i^\text{l},v_i^\text{r}\) and their neighbours, respectively.
    \item For constraints on holonomy angels on \(i\)-th loop: \(\sum_{j\in G_i}\mathbf{L}_{i,j}^\text{l}(\mathbf{u}+\mathbf{\delta u})=\frac{\pi}{2}r_i-\sum_{j\in G_i}k^g_{i,j}\).
\end{enumerate}

By designing the variables in this way and using \(\mathbf{\delta u}\) to express $\mathbf{u}$, we only introduce \(2g\) additional variables, which exactly match the number of added constraints.  
After assembling the above four parts into a complete linear system and moving all \(\mathbf{\delta u}\) terms to the right hand side, we get:
\begin{equation}
    \mathbf{L}_\text{g}\mathbf{u}=\frac{\pi}{2}
    \begin{pmatrix}
    \mathbf{z}\\
    \mathbf{r}
    \end{pmatrix}
    -
    \begin{pmatrix}
    \mathbf{k}^\text{ori}\\
    \mathbf{k}^g
    \end{pmatrix}
    +\mathbf{K}\mathbf{\delta u},
\end{equation}
where \(\mathbf{k}^g=(\sum_{j\in G_1}k^g_{1,j},\ldots, \sum_{j\in G_{2g}}k^g_{2g,j})^T\) and \(\mathbf{L}_\text{g},\mathbf{K}\) are \(N'\times N'\) matrices (\(N'=N+2g\)).
The rank of \(\mathbf{L}_\text{g}\) is \(N'-1\) \tb{(we discuss it in supplements)}, and \(\mathbf{L}_\text{g}\mathbf{1}_{N'}=\mathbf{0}\) holds, where \(\mathbf{1}_{N'}\) is the all-ones vector. Therefore, we follow the method mentioned in Sec.~\ref{subsubsec:solving_angles} and conclude that
\begin{equation}\label{reduction_equation_on_highgenus}
    \mathbf{u}=\widehat{\mathbf{L}}^{-1}\mathbf{M}\left(\frac{\pi}{2}
    \begin{pmatrix}
    \mathbf{z}\\
    \mathbf{r}
    \end{pmatrix}
    -
    \begin{pmatrix}
    \mathbf{k}^\text{ori}\\
    \mathbf{k}^g
    \end{pmatrix}
    +\mathbf{K}\mathbf{\delta u}\right)+a\mathbf{1}_{N'}.
\end{equation}
Therefore, in this way, instead of computing \(\text{Null}(\mathbf{L}_\text{h})\) numerically, we can explicitly assemble \(\mathbf{K}\) to construct the bases of the null space.

\begin{figure}[t]
  \centering
  \begin{overpic}[width=0.8\linewidth]{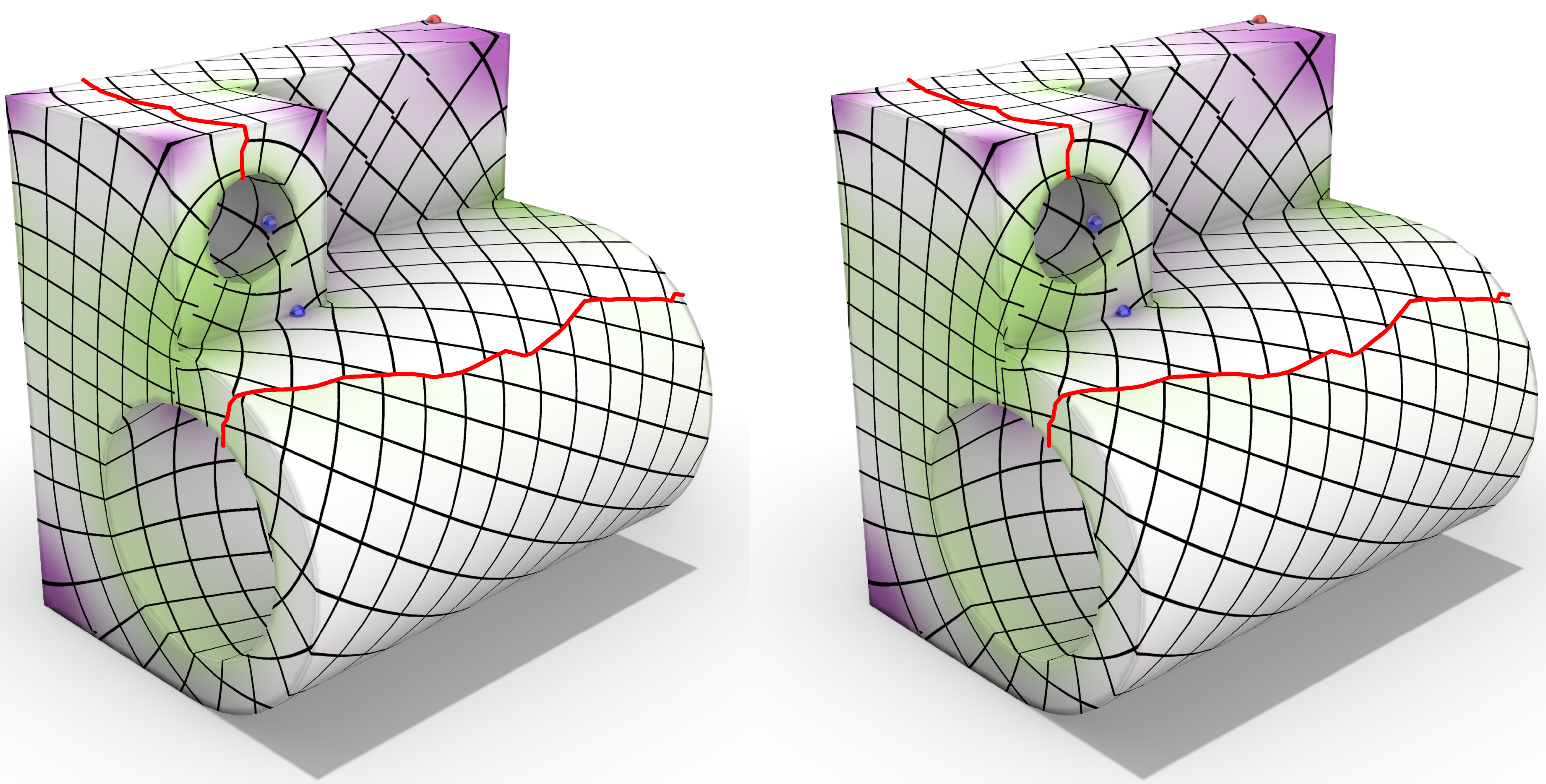}
    {
 \put(12,-4){\small 12.2 seconds} 
 \put(18,-9){\small \cite{zhang2023practical}}
 \put(66,-4){\small 2.9 seconds}
 \put(72.5,-9){\small Ours}
    }
  \end{overpic}
  \vspace{5mm}
  \caption{
  Comparison with~\cite{zhang2023practical} for solving the integer variables in~\eqref{discrete_holonomy_constraint}.
  With the same cone configuration and \(\lambda_d\), the same results are obtained by both methods, but ours is faster.
  }
  \label{fig:reduction_on_high-genus_surface}
\end{figure}

\paragraph{Formulation}
The energy term used to penalize the jump of \( \mathbf{u} \) across cuts can be written as  
\begin{equation}
\mathcal{E}_\text{dif}=\sqrt{\sum_{i=1}^{2g} \sum_{j \in G_i} \Vert u_j^\text{l} + \delta u_j^\text{l} - u_j^\text{r} - \delta u_j^\text{r} \Vert_2^2},
\end{equation}
where \(u_j^\text{l}\) and \(u_j^\text{r}\) are two opposite vertices on cuts.
By substituting~\eqref{reduction_equation_on_highgenus}, \(\mathbf{u}\) can be eliminated, and the expression is written as
\begin{equation}
\mathcal{E}_\text{dif}=\sqrt{\sum_{i=1}^{2g} \sum_{j \in G_i}\Vert \mathbf{F}_j(\mathbf{r},\mathbf{\delta u},a)\Vert_2^2},
\end{equation}
where \(\mathbf{F}_j\) is a linear vector-valued function with respect to \(\mathbf{r},\mathbf{\delta u},a\).
Hence, our formulation is:
\begin{equation}\label{small_scale_problem_in_highgenus}
\begin{aligned}
    \min_{\mathbf{r},\mathbf{\delta u},a}\ & \Vert\mathbf{A}^{\frac{1}{2}}(\frac{\pi}{2}\widehat{\mathbf{L}}^{-1}\mathbf{M}
\begin{pmatrix}
    \mathbf{0}\\
    \mathbf{r}
\end{pmatrix}
+\widehat{\mathbf{L}}^{-1}\mathbf{M}
\begin{pmatrix}
    \frac{\pi}{2}\mathbf{z}-\mathbf{k}^\text{ori}\\
    -\mathbf{k}^g
\end{pmatrix})
+a\mathbf{1}_{N'}\\&
+\widehat{\mathbf{L}}^{-1}\mathbf{M}\mathbf{K}\mathbf{\delta u}
\Vert_2^2
    +\lambda_d\sum_{i=1}^{2g} \sum_{j \in G_i}\Vert \mathbf{F}_j(\mathbf{r},\mathbf{\delta u},a)\Vert_2^2,\\
    s.t. \ &\mathbf{r}\in\mathbb{Z}^{2g},
\end{aligned}
\end{equation}
where \(\lambda_d\) is a parameter to adjust the difference of the log conformal factor on either side of cuts.
As illustrated in Fig.~\ref{fig:reduction_on_high-genus_surface}, our variable reduction method outperforms~\cite{zhang2023practical}, achieving faster computation.

\begin{figure}[t]
  \centering
  \begin{overpic}[width=0.99\linewidth]{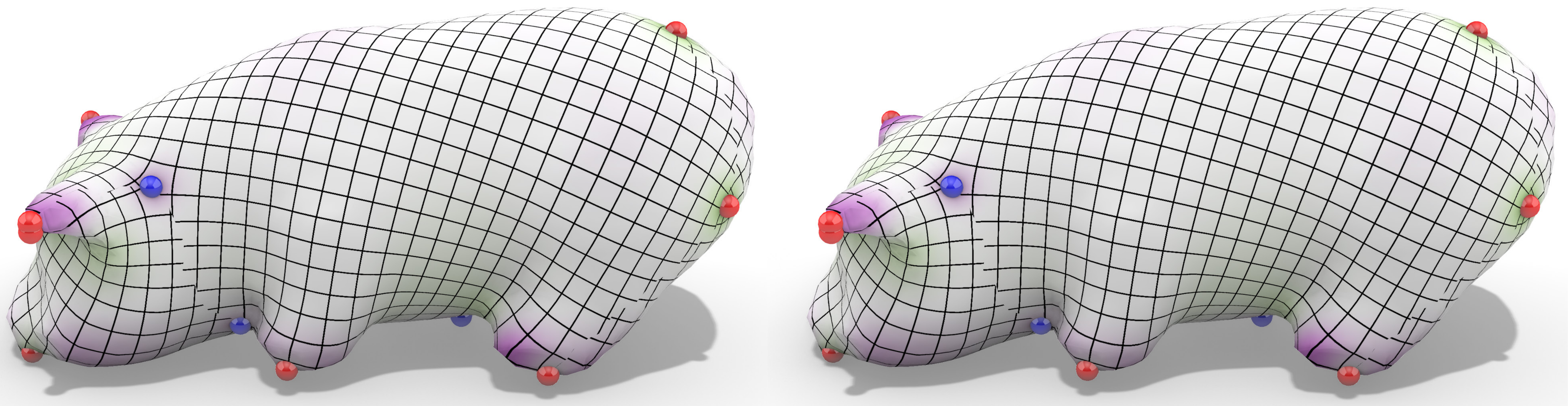}
    {
 \put(18,-4){\small 4.3 seconds}
 \put(69,-4){\small 1.8 seconds}
    }
  \end{overpic}
  \vspace{0mm}
  \caption{
  Solving angles after each cone movement (left) vs. Solving angles after cones converge (right).
  Both settings yield 14 cones with the same positions and angles, but the runtimes differ.
  }
  \label{fig:reducing_frequency_of_solving_angle}
\end{figure}


\section{Experiments and evaluations}\label{sec:results}
We implemented our algorithm in C++ and conducted all experiments
on a computer with an i7-11700 2.5GHz CPU and 16GB of RAM. \tb{The code has been released at XXX.com}. The integer optimizer used was Gurobi~\cite{gurobi}, while linear matrix factorization and solving were performed using Eigen~\cite{eigenweb}. For conformal parameterization, we utilized the Boundary First Flattening method (BFF)~\cite{Sawhney:2017:BFF}, and non-contractible homology loops were computed using the method described in~\cite{dey2013efficient}. For surfaces with boundaries, we adopt Dirichlet boundary conditions, that is, we set \(\mathbf{b}=\mathbf{0}\).

\tb{Unless otherwise specified, all our experiments are conducted under the same set of parameters.
The default parameters are set as target distortion \(\epsilon_\text{tar}=0.200\), the maximum number of integer variables in~\eqref{small_scale_problem} \(N_g = 30\) and weight in~\eqref{small_scale_problem_in_highgenus} \(\lambda_d=10^6\). 
The integer variables are by default restricted to \( N_b = [-1, 1] = \{-1, 0, 1\} \), 
which means that the cone angle can only take the values \(-\frac{\pi}{2}\), \(0\), or \(\frac{\pi}{2}\).
Below each example, we provide the number of generated cones \(N_c\), 
the final achieved distortion, and the computation time, similar to Fig.~\ref{fig:teaser}.}

\begin{figure}[t]
  \centering
  \begin{overpic}[width=0.99\linewidth]{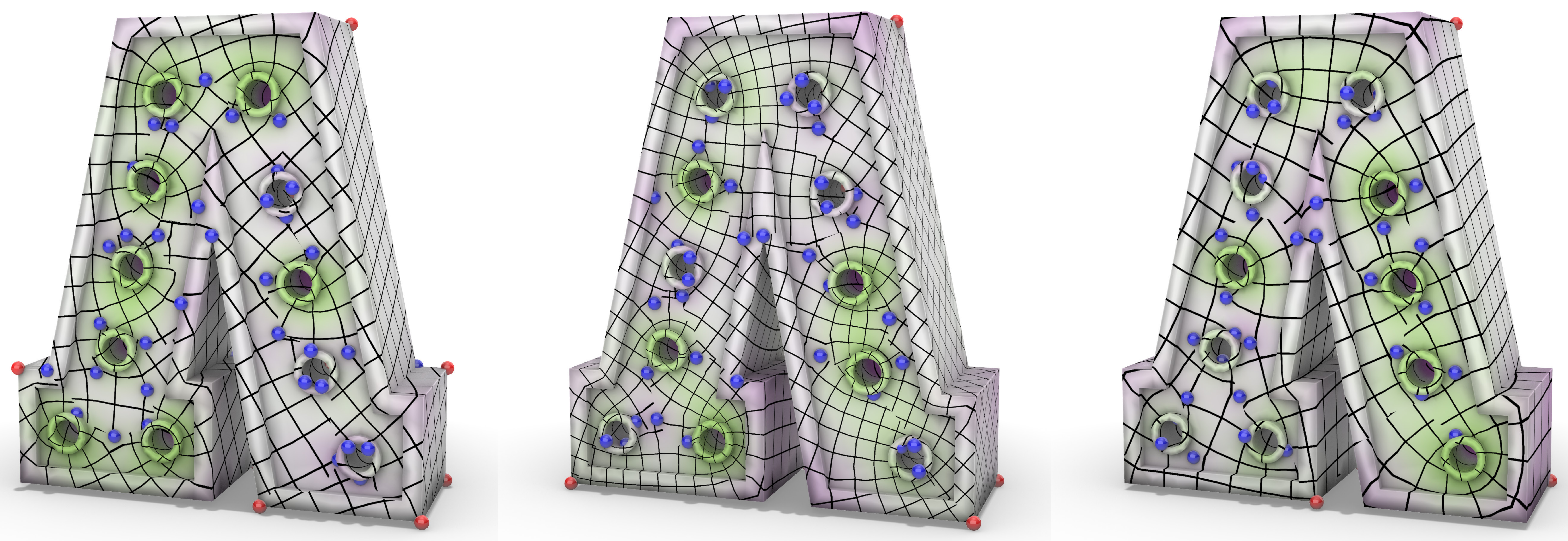}
    {
 \put(4,-3){\small (86, 0.203, 31.0)}
 \put(10,-7){\small \(N_g=10\)}
 \put(39,-3){\small (86, 0.192, 6.7)}
 \put(45,-7){\small \(N_g=30\)}
 \put(74,-3){\small (86, 0.187, 54.3)}
 \put(80,-7){\small \(N_g=\infty\)}
   }
  \end{overpic}
  \vspace{3mm}
  \caption{
  Active pruning strategy with different $N_g$s.
  }
  \label{fig:different_N_g}
\end{figure}

\begin{figure}[t]
  \centering
  \begin{overpic}[width=0.99\linewidth]{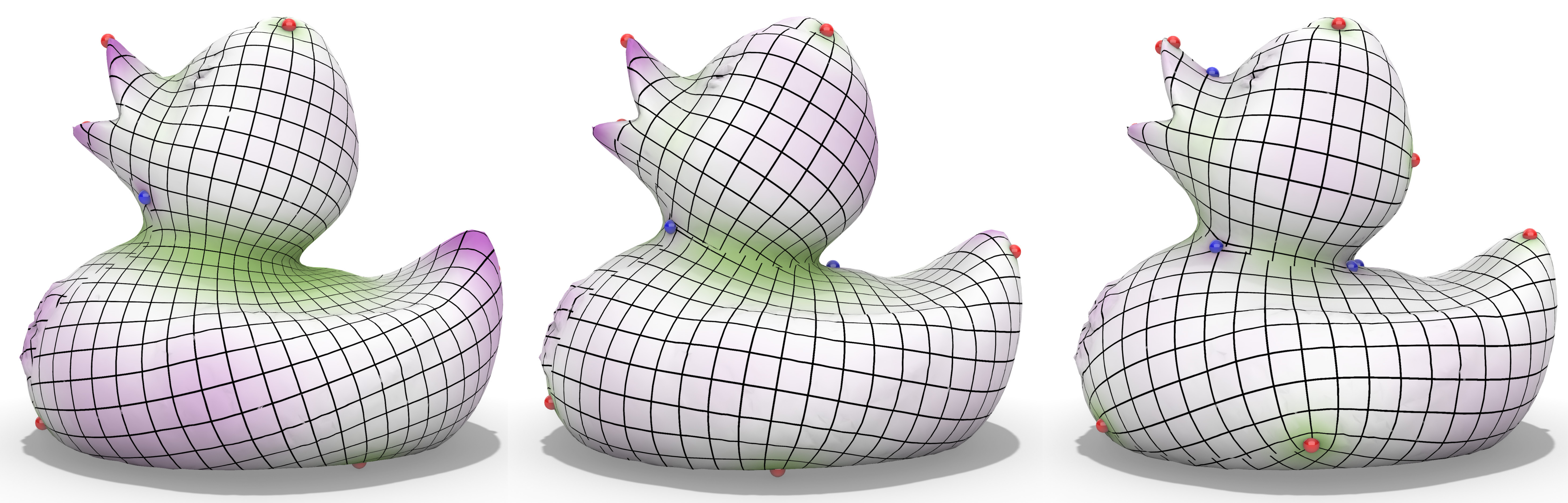}
    {
 \put(5,-3){\small (10, 0.187, 1.7)}
 \put(9,-7){\small \(\epsilon_\text{tar}=0.20\)}
 \put(38,-3){\small (12, 0.149, 2.6)}
 \put(42,-7){\small \(\epsilon_\text{tar}=0.15\)}
 \put(72,-3){\small (18, 0.099, 4.5)}
 \put(76,-7){\small \(\epsilon_\text{tar}=0.10\)}
    }
  \end{overpic}
   \vspace{4mm}
  \caption{
  Results of different \(\epsilon_\text{tar}\).
  }
  \label{fig:different_E_target}
\end{figure}

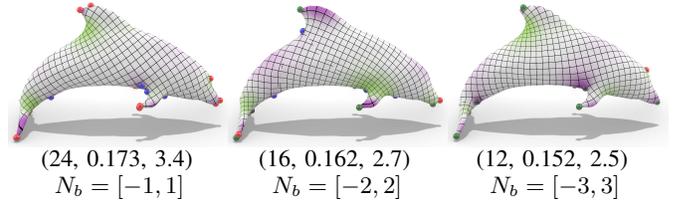
\begin{figure}[t]
  \centering
  \begin{overpic}[width=0.99\linewidth]{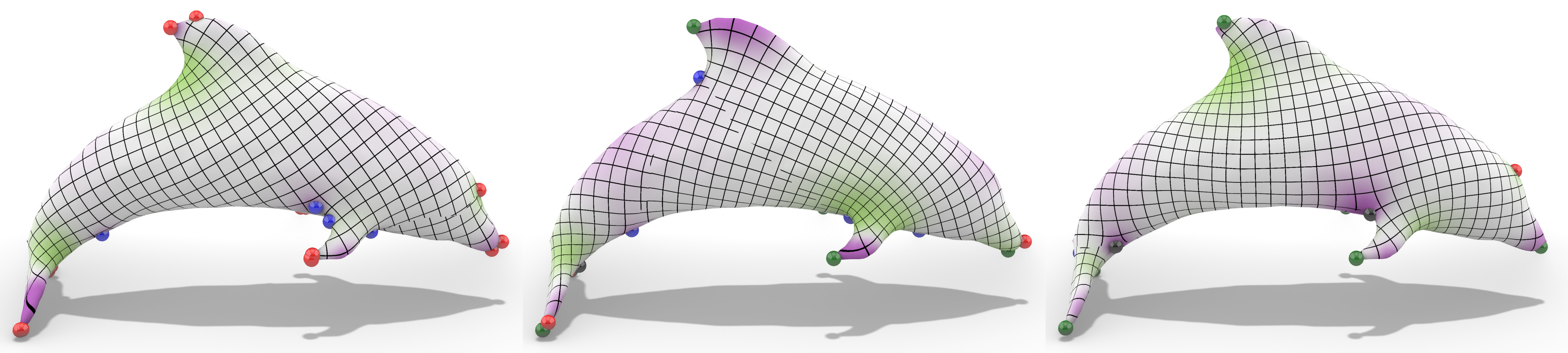}
    {
 \put(5,-3){\small (24, 0.173, 3.4)}
 \put(7,-7){\small \(N_b=[-1,1]\)}
 \put(38,-3){\small (16, 0.162, 2.7)}
 \put(40,-7){\small \(N_b=[-2,2]\)}
 \put(71,-3){\small (12, 0.152, 2.5)}
 \put(73,-7){\small \(N_b=[-3,3]\)}
    }
  \end{overpic}
  \vspace{4mm}
  \caption{
  Results of different \(N_b\). The input distortion \(\epsilon_\text{tar}\) are set to 0.200, 0.173 and 0.162, where the input distortion of the last two is the output distortion of the previous.
  }
  \label{fig:different_integer_bounds}
\end{figure}

\subsection{Intra-algorithm evaluations}
\paragraph{Solving angles after each cone movement}
Solving angles after each cone movement may accelerate the reduction of distortion, which is adopted by~\cite{li2023efficient}.
We choose to move the cones until the distortion converges without updating angles. 
These two strategies are compared in Fig.~\ref{fig:reducing_frequency_of_solving_angle}.
Our scheme is much faster and achieves identical results.
We observe that the cone angles remain almost unchanged during cone movement; thus, our strategy is effective.
Another reason is that the cost of solving angles is much higher than that of moving cones.


\paragraph{Active pruning strategy}
The number of feasible integer combinations to (\ref{small_scale_problem}) is \(|N_b|^{N_g-1}\), \tb{indicating that the number of integer variables \(N_g\) (the number of active cones) has a significant impact on the efficiency of solving (\ref{small_scale_problem}).}
We evaluate how different \(N_g\)s affect both the optimization efficiency and the solution quality in Fig.~\ref{fig:different_N_g}. 
If \( N_g \) is too small, the pruned solution space becomes too restricted, and the target distortion cannot be reached after the maximum number of iterations. 
On the other hand, if no constraint is imposed on \( N_g \), lower distortion can be achieved, but at the cost of significantly increased computation time.
In our experiments, setting \( N_g \) in the range of 20 to 60 provides a good balance between distortion and computation time for most cases; we use 30 in practice.

\paragraph{Different parameters}
We test our algorithm under different target distortions \(\epsilon_\text{tar}\) in Fig.~\ref{fig:different_E_target}. As \(\epsilon_\text{tar}\) decreases, both the number of cones and the computation time increase.

Our approach remains effective over a broader range of integer values, as demonstrated in Fig.~\ref{fig:different_integer_bounds}. For instance, on the Dolphin's fin, cone singularities of $\pi$ or even $3\pi/2$ can be accommodated. Expanding the allowable range of integers reduces the number of singularities required to achieve a comparable level of distortion.

\paragraph{Different meshes}
Next, we present the results of our algorithm on various meshes to demonstrate its adaptability.



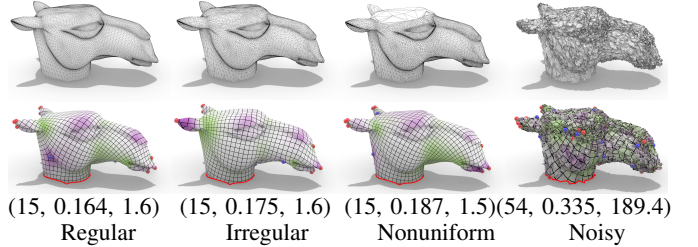
\begin{figure}[t]
  \centering
  \begin{overpic}[width=0.99\linewidth]{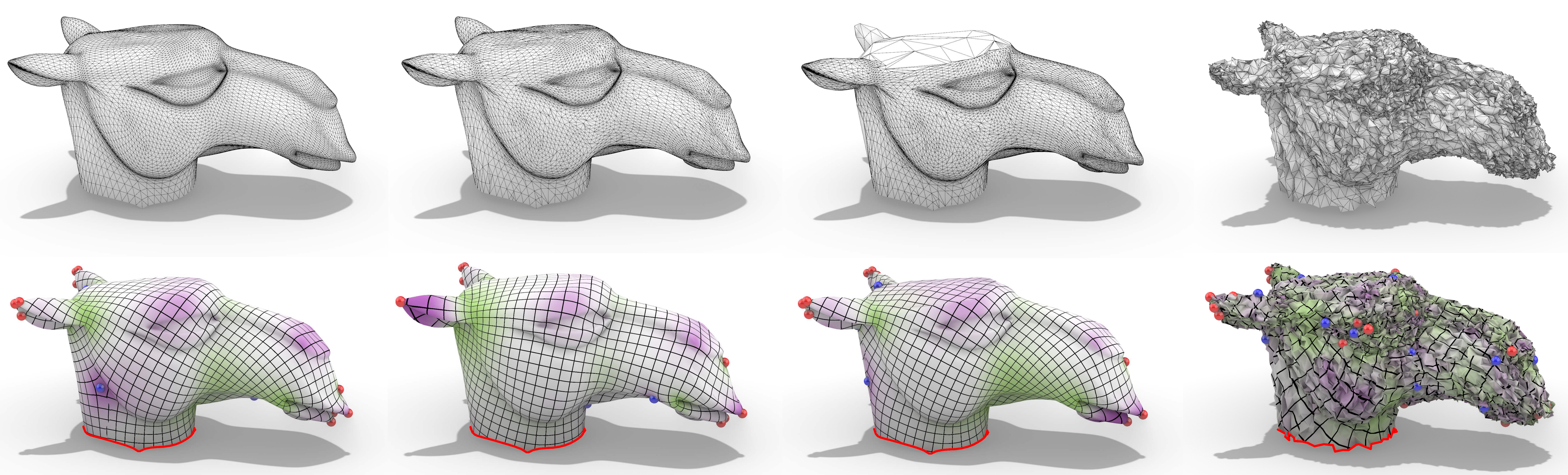}
    {
 \put(0,-3){\small (15, 0.164, 1.6)}
 \put(26,-3){\small (15, 0.175, 1.6)}
 \put(51,-3){\small (15, 0.187, 1.5)}
 \put(74,-3){\small (54, 0.335, 189.4)}
 \put(8,-7){\small Regular}
 \put(33,-7){\small Irregular}
 \put(56,-7){\small Nonuniform}
 \put(85,-7){\small Noisy}
    }
  \end{overpic}
  \vspace{3mm}
  \caption{
  Results of different tessellations on camel model.
  }
  \label{fig:different_triangulations}
\end{figure}

We evaluate our algorithm on the Camel model under four different types of tessellations (regular, irregular, nonuniform, and noisy) in Fig.~\ref{fig:different_triangulations}, provided by~\cite{Soliman2018OCS}.
For the regular, irregular, and nonuniform tessellations, the results are similar. 
In the noisy tessellation, the increased number of vertices with high curvature necessitates additional cones to adequately reduce the distortion.
Nonetheless, there exists a lower bound on the achievable distortion, and our algorithm is unable to reduce it to 0.200.
This reveals a fundamental difference between conformal parameterization with integer constraints and that without such constraints~\cite{fang2021computing,li2023efficient}.

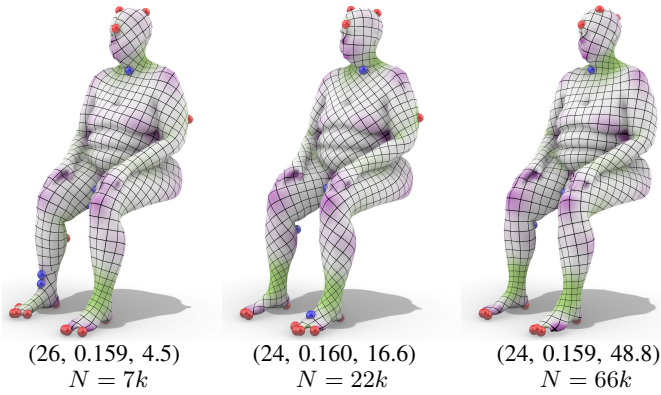
\begin{figure}[t]
  \centering
  \begin{overpic}[width=0.99\linewidth]{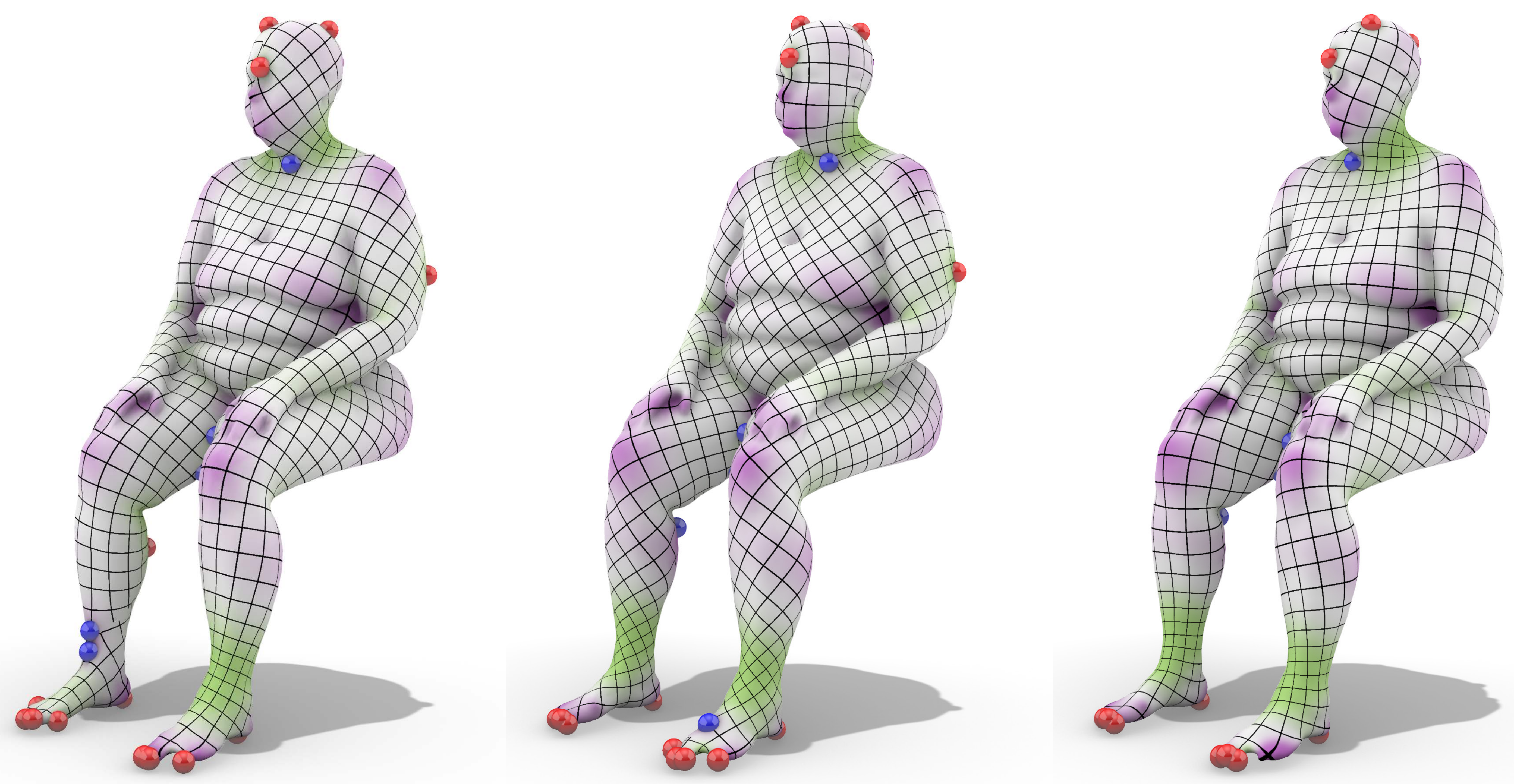}
    {
 \put(4,-3){\small (26, 0.159, 4.5)}
 \put(10,-7){\small \(N=7k\)}
 \put(38,-3){\small (24, 0.160, 16.6)}
 \put(44.5,-7){\small \(N=22k\)}
 \put(75,-3){\small (24, 0.159, 48.8)}
 \put(81.5,-7){\small \(N=66k\)}
    }
  \end{overpic}
  \vspace{3mm}
  \caption{
  Results of different numbers of vertices on the Human model. The input distortions \(\epsilon_\text{tar}\) are set to 0.166, 0.163, and 0.170 to achieve a similar output distortion.
  }
  \label{fig:different_magnitudes}
\end{figure}

We test the Human model (Fig.~\ref{fig:different_magnitudes}) with three different numbers of vertices. 
The resulting cone number \(N_c\) and distortion \(\mathcal{E}\) are similar, and the time concerning the number of vertices is approximately linear.

\begin{figure}[t]
  \centering
  \begin{overpic}[width=0.99\linewidth]{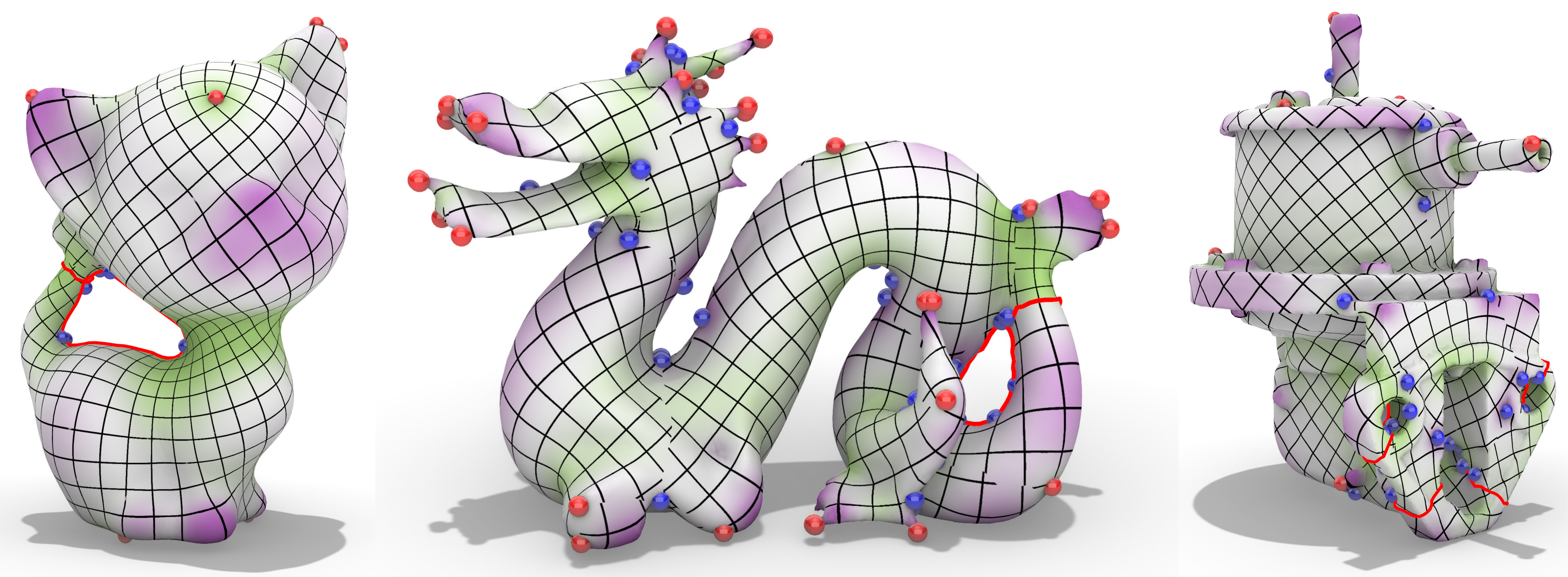}
    {
 \put(0,-3.5){\small (14, 0.177, 1.8)}
 \put(36,-3.5){\small (82, 0.190, 38.4)}
 \put(75,-3.5){\small (62, 0.191, 27.6)}
    }
  \end{overpic}
  \vspace{1mm}
  \caption{
  Nonzero genus surfaces with genus 1, 1, and 4, respectively. The red paths indicating the cut paths. The input distortion \(\epsilon_\text{tar}\) are set to 0.180, 0.185, 0.190 to ensure the resulting distortion is below 0.200.
  }
  \label{fig:high-genus surfaces}
\end{figure}

Our algorithm is capable of handling nonzero genus meshes, and the rotationally seamless conformal parameterizations generated exhibit small distortion differences across the cuts. Fig.~\ref{fig:high-genus surfaces} shows three models with nonzero genus.

\paragraph{Dataset statistics}
Similar to~\cite{Li2022int,zhang2023practical}, we test our algorithm on the dataset containing 3885 mechanical and organic models provided by~\cite{ZhuGreedy2020} to evaluate its effectiveness, robustness, and efficiency. The average \textblue{number} of the mesh vertices in the dataset is \(6.8k\).

\begin{figure}[t]
  \centering
  \begin{overpic}[width=0.99\linewidth]{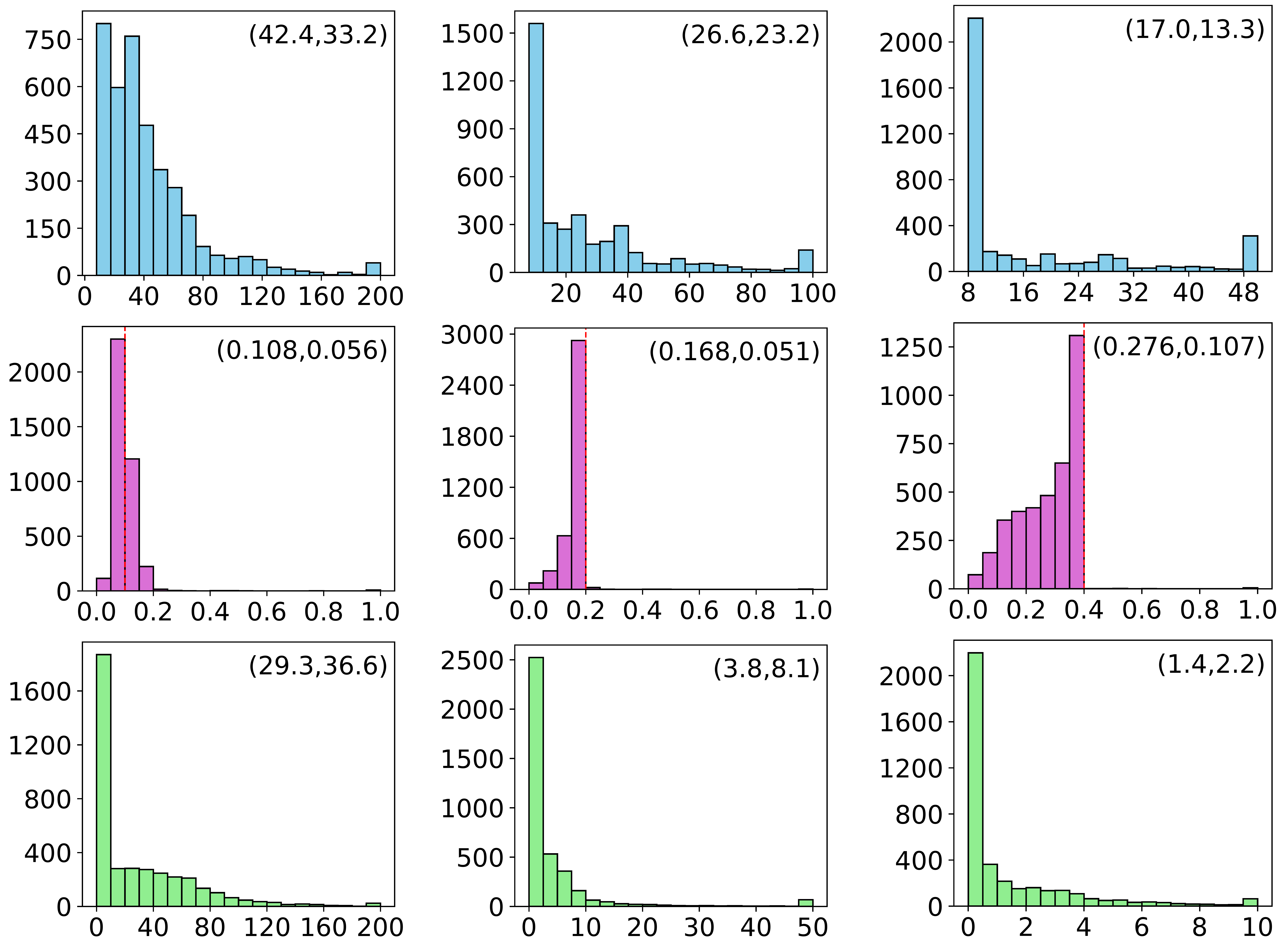}
    {
 \put(12,-3){\small \(\epsilon_\text{tar}=0.1\)}
 \put(45,-3){\small \(\epsilon_\text{tar}=0.2\)}
 \put(80,-3){\small \(\epsilon_\text{tar}=0.4\)}
    }
  \end{overpic}
  \vspace{1mm}
  \caption{
  The distributions of the number of vertices \(N_c\) (upper), the resulting distortion \(\mathcal{E}\) (middle), and the runtime \(t\) (bottom, in seconds) via histograms for dataset testing. The average and standard deviation are shown in the right-upper side of each histogram.
  }
  \label{fig:dataset_statistics}
\end{figure}

Three different target distortions are used, including \(\epsilon=0.1, 0.2\) and \(0.4\). 
We collect statistics on the number of cones \(N_c\), distortion \(\mathcal{E}\), and computation time \(t\) (in seconds), as shown in Fig.~\ref{fig:dataset_statistics}.
Except for the group of \(\epsilon=0.1\), our algorithm typically reaches the target distortion bound within 10 seconds.
The success rates of reaching the target distortion \textblue{in prescribed number of iterations (we set it to 1000)} in the three groups are 62.16\%, 99.10\% and 99.74\%, respectively.

\paragraph{Local injectivity}\label{para:local_injectivity}
\tb{We focus on efficiently computing sparse cones, rather than computing locally injective parameterizations. Theoretically, conformal parameterizations should not exhibit flips.
We adopt the method of~\cite{Sawhney:2017:BFF} for parameterizations where injectivity is not guaranteed. In our dataset (Fig.~\ref{fig:dataset_statistics}), when the target distortions are set to 0.1, 0.2, and 0.4, the percentages of parameterizations in the results that contain flipped triangles are 22.1\%, 19.4\%, and 18.8\%, respectively. Within each case, the proportion of flipped triangles is 0.2\%.
However, if alternative parameterization methods are used, e.g.,~\cite{capouellez2022metric}, local injectivity is guaranteed.
}

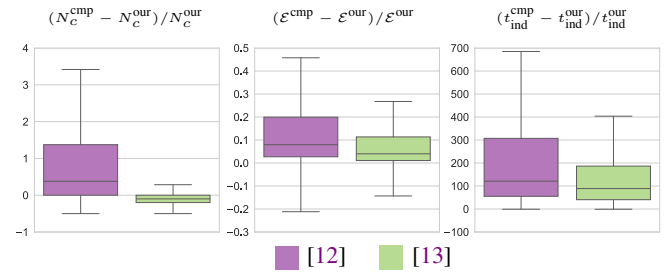
\begin{figure}[t]
  \centering
  \vspace{4mm}
  \begin{overpic}[width=0.99\linewidth]{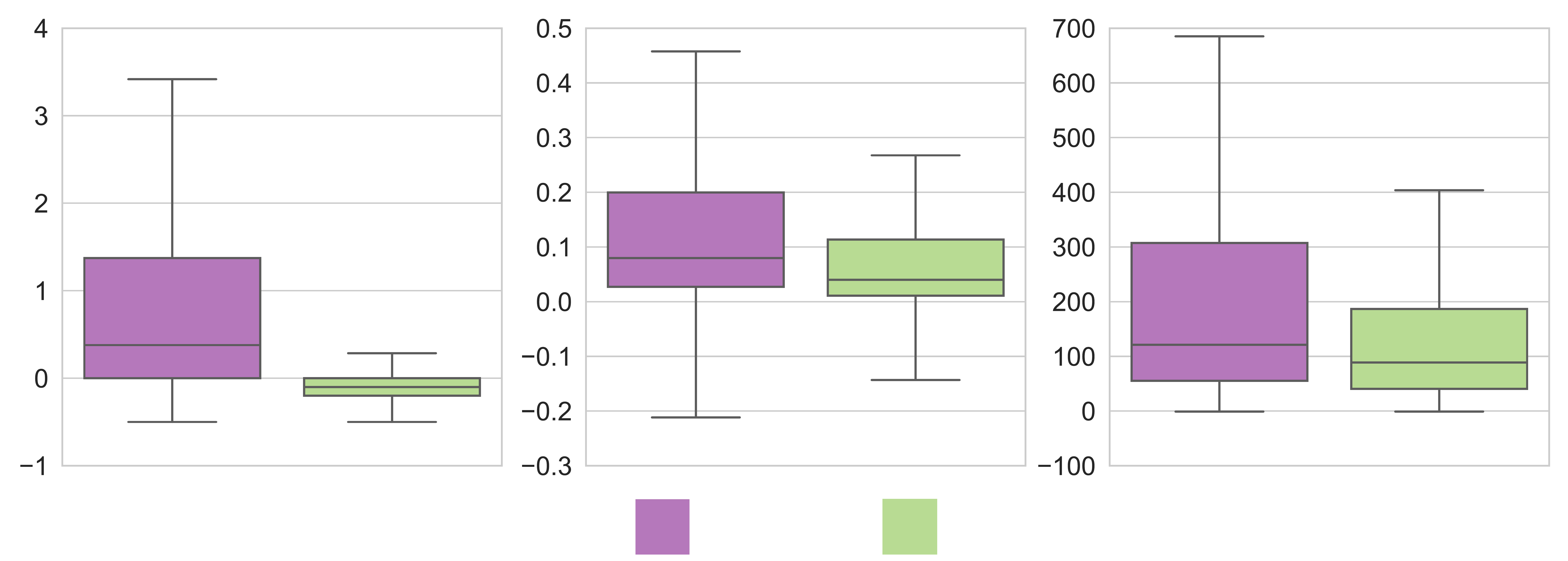}
    {
 \put(6,38){\tiny \((N_c^\text{cmp}-N_c^\text{our})/N_c^\text{our}\)}
 \put(40,38){\tiny \((\mathcal{E}^\text{cmp}-\mathcal{E}^\text{our})/\mathcal{E}^\text{our}\)}
 \put(74,38){\tiny \((t_\text{ind}^\text{cmp}-t_\text{ind}^\text{our})/t_\text{ind}^\text{our}\)}
 \put(44,1.3){\small \cite{Li2022int}}
 \put(60,1.3){\small \cite{zhang2023practical}}
    }
  \end{overpic}
  \vspace{-3mm}
  \caption{
  Box plots for \((\text{X}^\text{cmp}-\text{X}^\text{our})/\text{X}^\text{our}\), where \(\text{X}\) denotes \(N_c\), \(\mathcal{E}\) and \(t_\text{ind}\).
  Here, \(t_\text{ind}^\text{cmp}\) and \(t_\text{ind}^\text{our}\) are the runtimes for each individual model processed by the competitor's and our algorithm.
  }
  \label{fig:cmp_statistics}
\end{figure}

\begin{figure}[t]
  \centering
  \begin{overpic}[width=0.99\linewidth]{gege}
    {
 \put(0,62){\small (15, 0.185, 85.0)}
 \put(25,62){\small (11, 0.169, 0.1)}
 \put(50,62){\small (8, 0.171, 21.7)}
 \put(75,62){\small (8, 0.158, 0.2)}
 
 \put(-1,30){\small (33, 0.230, 460.2)}
 \put(26,30){\small (24, 0.225, 3.8)}
 \put(50,30){\small (42, 0.167, 238.1)}
 \put(76,30){\small (42, 0.157, 10.2)}
 
 \put(-1,-4){\small (43, 0.248, 404.5)}
 \put(25,-4){\small (27, 0.197, 4.2)}
 \put(49,-4){\small (46, 0.155, 1364.3)}
 \put(77,-4){\small (42, 0.141, 10.8)}

 \put(8,-8.5){\small \cite{Li2022int}}
 \put(33,-8.5){\small Ours}
 \put(58,-8.5){\small \cite{zhang2023practical}}
 \put(84,-8.5){\small Ours}
    }
  \end{overpic}
  \vspace{5mm}
  \caption{
  Models for \(N_c^\text{cmp} \ge N_c^\text{our}\) and \(\mathcal{E}^\text{cmp}> \mathcal{E}^\text{our}\).
  }
  \label{fig:gege}
\end{figure}

\begin{figure}[t]
  \centering
  \begin{overpic}[width=0.99\linewidth]{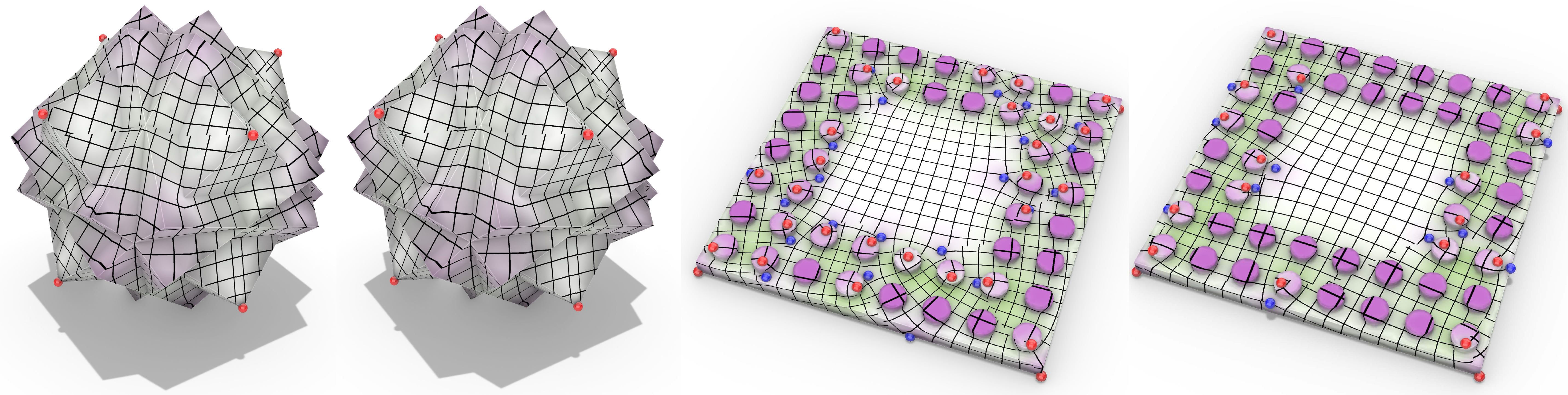}
    {
 \put(0,-4){\small (8, 0.168, 4.7)}
 \put(22,-4){\small (8, 0.168, 0.2)}
 \put(45,-4){\small (34, 0.183, 370.1)}
 \put(74,-4){\small (32, 0.195, 92.1)}
 
 \put(7,-8){\small \cite{Li2022int}}
 \put(29,-8){\small Ours}
 \put(54,-8){\small \cite{zhang2023practical}}
 \put(83,-8){\small Ours}
    }
  \end{overpic}
  \vspace{4mm}
  \caption{
  Models for \(N_c^\text{cmp} \ge N_c^\text{our}\) and \(\mathcal{E}^\text{cmp}\le \mathcal{E}^\text{our}\).
  }
  \label{fig:gele}
\end{figure}

\begin{figure}[t]
  \centering
  \begin{overpic}[width=0.99\linewidth]{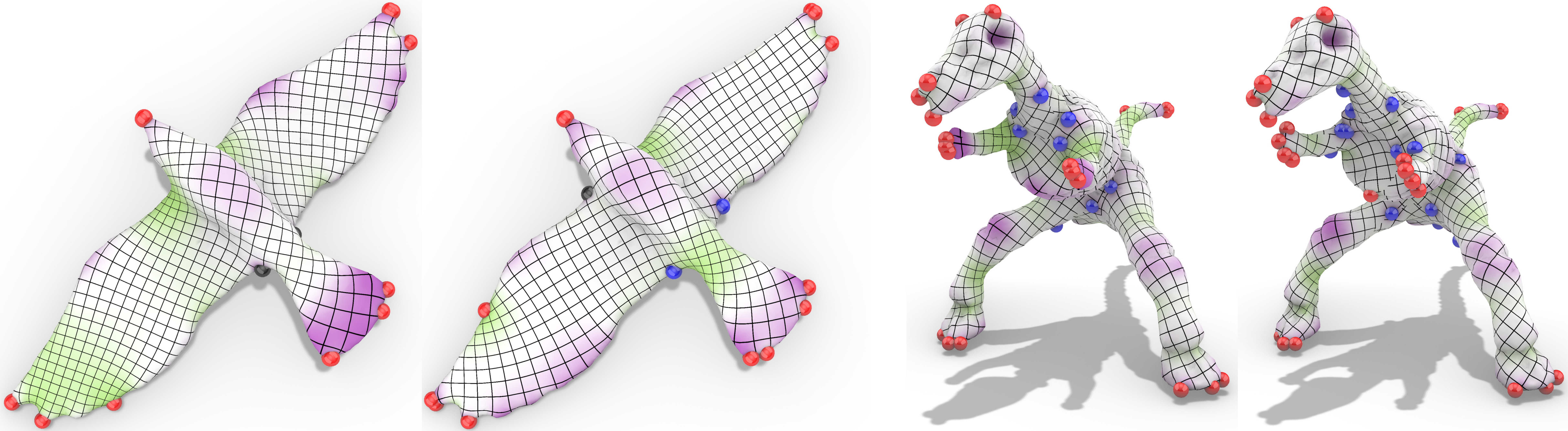}
    {
 \put(0,-4){\small (17, 0.202, 255.5)}
 \put(28,-4){\small (20, 0.162, 3.1)}
 \put(53,-4){\small (38, 0.181, 241.4)}
 \put(79,-4){\small (44, 0.152, 4.3)}
 
 \put(10,-8){\small \cite{Li2022int}}
 \put(38,-8){\small Ours}
 \put(62,-8){\small \cite{zhang2023practical}}
 \put(86,-8){\small Ours}
    }
  \end{overpic}
  \vspace{4mm}
  \caption{
  Models for \(N_c^\text{cmp} < N_c^\text{our}\) and \(\mathcal{E}^\text{cmp}> \mathcal{E}^\text{our}\).
  }
  \label{fig:lege}
\end{figure}

\subsection{Comparisons}
\paragraph{Competitors}
We compare with~\cite{Li2022int} and~\cite{zhang2023practical} on the dataset for generating integer-constrained cones. 
We use the distortion outputs of these two methods as the target distortion \(\epsilon_{tar}\)s for our algorithm.
Since \cite{Li2022int} and \cite{zhang2023practical} adopt different integer bounds, we ensure fairness by adopting the same bounds when comparing with each of them.
The number of cones produced by each method is denoted as \(N_c^\text{cmp}\) for the competitors and \(N_c^\text{our}\) for ours. 
Similarly, the resulting distortion for the competitor and ours are denoted as \(\mathcal{E}^\text{cmp}, \mathcal{E}^\text{our}\), respectively.

\paragraph{Results}
In Fig.~\ref{fig:cmp_statistics},  \((\text{X}^\text{cmp}-\text{X}^\text{our})/\text{X}^\text{our}\) are illustrated, where \(\text{X}\) denotes \(N_c\), \(\mathcal{E}\) and \(t_\text{ind}\).
Here, \(t_\text{ind}^\text{cmp}\) and \(t_\text{ind}^\text{our}\) are the runtimes for each individual model processed by the competitor's and our algorithm.

\begin{figure}[t]
  \centering
  \begin{overpic}[width=0.99\linewidth]{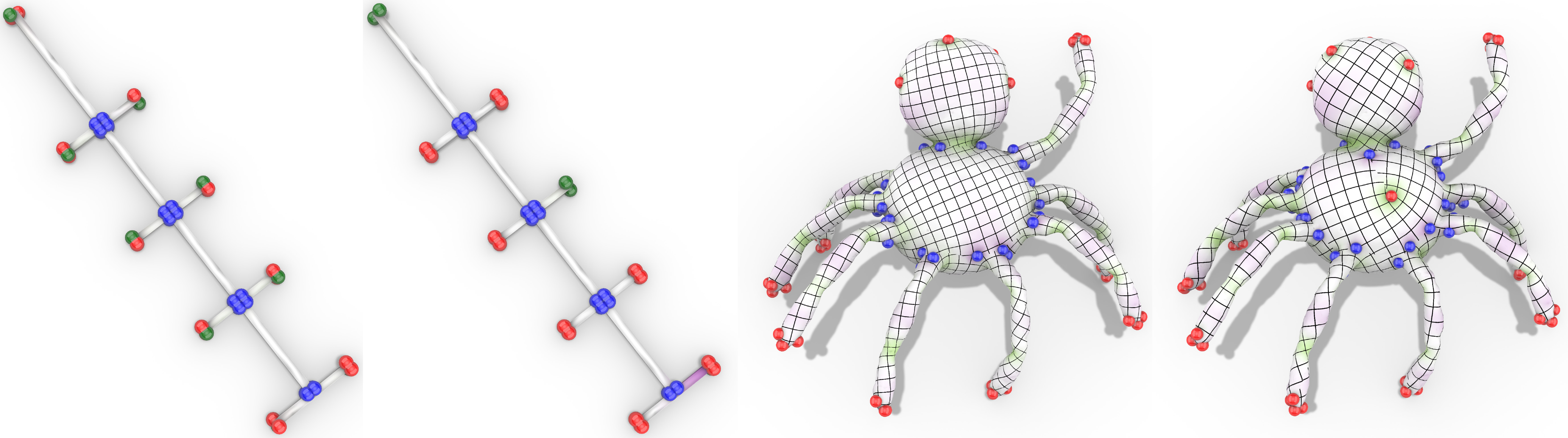}
    {
 \put(0,-4){\small (55, 0.094, 1.9)}
 \put(23,-4){\small (58, 0.203, 10.0)}
 \put(48,-4){\small (66, 0.129, 254.7)}
 \put(75,-4){\small (68, 0.131, 116.0)}

 \put(6,-8){\small \cite{Li2022int}}
 \put(32,-8){\small Ours}
 \put(56,-8){\small \cite{zhang2023practical}}
 \put(84,-8){\small Ours}
    }
  \end{overpic}
  \vspace{4mm}
  \caption{
  Models for \(N_c^\text{cmp} < N_c^\text{our}\) and \(\mathcal{E}^\text{cmp}\le \mathcal{E}^\text{our}\).
  }
  \label{fig:lele}
\end{figure}

To further investigate the statistics, we categorize the results based on the relationship between cone count and distortion.
We also record the total computation time for each category of results, denoted as \(t^\text{cmp}\) and \(t^\text{our}\), respectively.
In Tab.~\ref{tab:lm},~\ref{tab:zz}, we report the number and proportion of results in each category, as well as for all results combined. 
Additionally, we present the averages and standard deviations of \(N_c^\text{cmp}-N_c^\text{our}\) and \(\mathcal{E}^\text{cmp}-\mathcal{E}^\text{our}\).
For the four categories, we present some representative models in Figs.~\ref{fig:gege},~\ref{fig:gele},~\ref{fig:lege},~\ref{fig:lele}, respectively.

In terms of effectiveness, compared with \cite{Li2022int}, our method produces significantly fewer cones, with 87.1\% of the cases achieving both fewer cones and lower distortion. 
Compared with \cite{zhang2023practical}, our method yields similar cone counts and distortion, with slightly better averages in both metrics.  
Regarding computational efficiency, our method achieves a substantial speed advantage over both methods, outperforming them by more than 30 times.

\subsection{Application}\label{subsec:application}
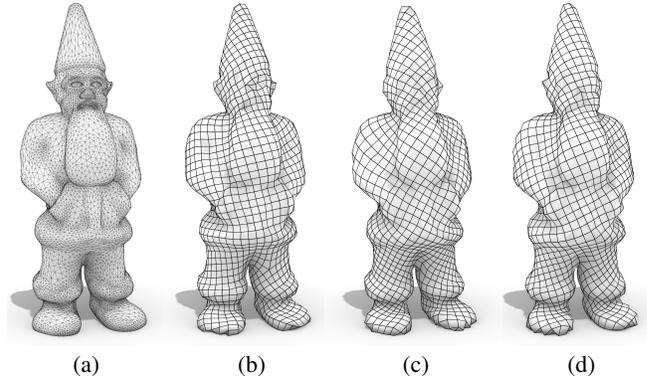
\begin{figure}[t]
  \centering
  \begin{overpic}[width=0.99\linewidth]{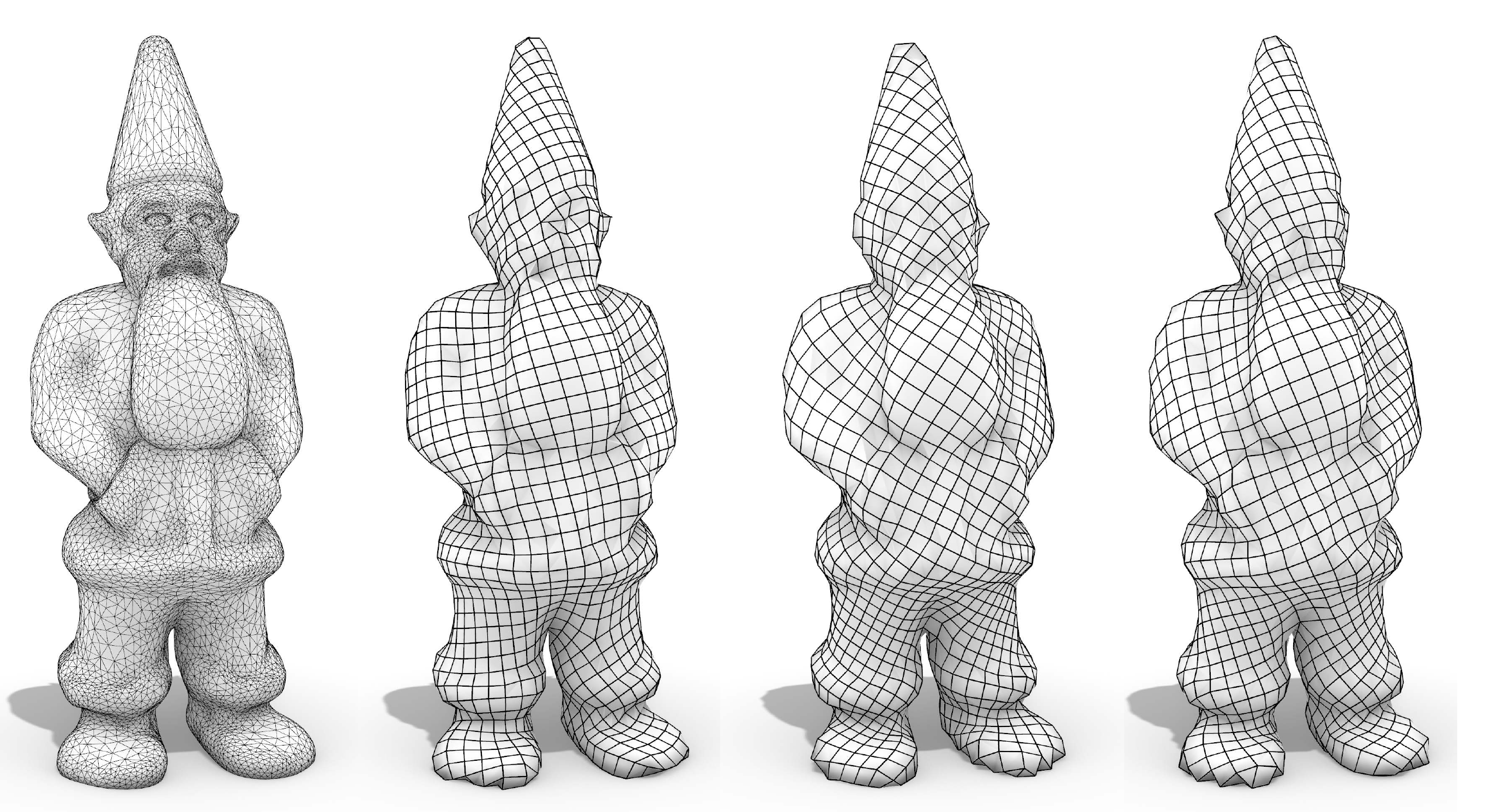}
    {
 \put(10,-4){\small (a)}
 \put(35,-4){\small (b)}
 \put(60,-4){\small (c)}
 \put(85,-4){\small (d)}
    }
  \end{overpic}
  \vspace{0mm}
  \caption{\tb{The input triangular mesh (a) is first parameterized using IOQ + libigl, \cite{zhang2023practical} + BFF, and ours + BFF, respectively.
Each parameterization is then followed by libQEx, which generates the corresponding quad meshes (b–d).}}
  \label{fig:quad_mesher}
\end{figure}

\tb{Our method can produce a seamless parameterization using BFF~\cite{Sawhney:2017:BFF}, followed by quad mesh extraction through libQEx~\cite{ebke2013qex}. 
For comparison, we also generate quad meshes using two alternative approaches.
In the first approach, we replace our cone generation method with that of \cite{zhang2023practical}; the second approach is to apply the pipeline of obtaining sparse cones (or integer-constrained cross field) using IOQ~\cite{farchi2018integer}, followed by a seamless parameterization via libigl~\cite{libigl} and then generating a quad mesh using libQEx, as shown in Fig.~\ref{fig:quad_mesher}.}


\tb{We record the statistics of the quadrilateral meshes generated by different methods, as shown in Tab.~\ref{tab:quad_eval}.
Since our method is based on conformal parameterization, the interior angles are more uniform, although the edge lengths exhibit relatively larger variations. 
In terms of computational efficiency, our method demonstrates a significant advantage.}

\section{Conclusion and Discussion}\label{sec:conclusion}
We propose an efficient method for generating sparse cone singularities, which can be used to derive low-distortion rotationally seamless conformal parameterizations. 
We iteratively optimize the cone number, angles, and positions in sequence to achieve this goal. 
When computing cone angles and holonomy angles for the non-contractible homology loops, we avoid the complexity of the original optimization by explicitly constructing an equivalent small-scale problem, which significantly improves efficiency. 
The efficiency and practical robustness of our algorithm are validated on more than 3885 models.

\paragraph{Global seamless conformal parameterizations}
Even on nonzero genus meshes, a globally seamless conformal parameterization exists if the Abel-Jacobi condition is satisfied~\cite{lei2020quadrilateral,zheng2021quadrilateral}. However, since our cone positions are restricted to mesh vertices, this condition typically cannot be fulfilled, leading to the general nonexistence of globally seamless conformal parameterizations. One potential solution is to dynamically refine the mesh by introducing new vertices, thereby enabling cone placement at more flexible locations and facilitating the construction of a globally seamless conformal parameterization. 

\paragraph{Limitations}
The main limitation of our method arises when the target distortion is set very low (e.g., 0.1), which often leads to a large number of cones and significantly slows down the optimization. 
A potential remedy is to design more effective pruning strategies. 
Additionally, the existing branch-driven and greedy cone insertion strategies are less effective for CAD models. This is because CAD surfaces typically contain numerous vertices with Gaussian curvature close to \(\pm\pi/2\). Directly identifying and assigning cones at these locations could lead to higher efficiency and lower distortion.
Finally, due to the discrete and combinatorial characteristics, we cannot theoretically guarantee that the final distortion is less than the specified threshold.

\paragraph{Discussion on applications}\label{para:more_discussion}
\tb{Conformal parameterization has its inherent limitations for wide range of applications. On the one hand, conformal methods are theoretically restricted and cannot achieve seamless parameterization if cones are restricted to vertices; 
on the other hand, imposing alignment constraints also presents theoretical challenges.}

\tb{However, we still consider our research promising, because we can also serve as a tool to generate cones for other types of parameterizations.
In geometry processing, the cone configuration has a significant impact on the parameterization result. However, most existing methods rely on simple heuristic strategies. 
\cite{myles2012global} first proposes using low–area-distortion and strict-conformal parameterization as a proxy for low-distortion isometric parameterization, 
thereby enabling direct optimization of the cone configuration. 
Subsequent studies, including ours, continue to improve the practical effectiveness of this idea. 
Hence, our algorithm can be regarded as a novel approach for solving the cone configuration problem, which can be directly followed by an isometric parameterization stage. 
Moreover, if strict conformality is not required, various existing methods~\cite{Campen2019Seamless,myles2012global,campen2017similarity} to obtain seamless parameterizations can be employed; 
if alignment constraints are needed, method decribed in~\cite{capouellez2025feature} can achieve this. 
Thus, we believe that our algorithm remains practically useful as a foundation for further parameterization tasks.}

\begin{table}[t]
\centering
\caption{
Comparison with \cite{Li2022int}.
}
\vspace{-2mm}
\small
\resizebox{1.0\linewidth}{!}{
\begin{tabular}{cccccc}
\toprule
Category & Count & Ratio & $N_c^\text{cmp} - N_c^\text{our}$ & $\mathcal{E}^\text{cmp} - \mathcal{E}^\text{our}$ & $\frac{t^\text{cmp}}{t^\text{our}}$ \\
\midrule
$N_c^\text{cmp} \ge N_c^\text{our} \wedge \mathcal{E}^\text{cmp} > \mathcal{E}^\text{our}$ & 3382 & 87.1\% & (50.7, 301.4) & (0.023, 0.034) & 50.1 \\
$N_c^\text{cmp} \ge N_c^\text{our} \wedge \mathcal{E}^\text{cmp} \le \mathcal{E}^\text{our}$ & 36 & 0.9\% & (388.9, 806.9) & (-0.575, 1.583) & 1.2 \\
$N_c^\text{cmp} < N_c^\text{our} \wedge \mathcal{E}^\text{cmp} > \mathcal{E}^\text{our}$ & 463 & 11.9\% & (-2.0, 1.4) & (0.037, 0.034) & 47.5 \\
$N_c^\text{cmp} < N_c^\text{our} \wedge \mathcal{E}^\text{cmp} \le \mathcal{E}^\text{our}$ & 2 & 0.1\% & (-60.5, 57.5) & (-0.235, 0.126) & 0.9 \\
$\text{Total}$ & 3883 & 100.0\% & (47.5, 294.2) & (0.019, 0.167) & 36.0 \\
\bottomrule
\end{tabular}
\label{tab:lm}
}
\end{table}

\begin{table}[!htbp]
\centering
\caption{Comparison with \cite{zhang2023practical}}
\vspace{-2mm}
\small
\resizebox{1.0\linewidth}{!}{
\begin{tabular}{cccccc}
\toprule
Category & Count & Ratio & $N_c^\text{cmp} - N_c^\text{our}$ & $\mathcal{E}^\text{cmp} - \mathcal{E}^\text{our}$ & $\frac{t^\text{cmp}}{t^\text{our}}$ \\
\midrule
$N_c^\text{cmp} \ge N_c^\text{our} \wedge \mathcal{E}^\text{cmp} > \mathcal{E}^\text{our}$ & 906 & 23.3\% & (10.0, 218.7) & (0.018, 0.347) & 105.5 \\
$N_c^\text{cmp} \ge N_c^\text{our} \wedge \mathcal{E}^\text{cmp} \le \mathcal{E}^\text{our}$ & 103 & 2.7\% & (74.1, 344.9) & (-0.253, 1.316) & 9.2 \\
$N_c^\text{cmp} < N_c^\text{our} \wedge \mathcal{E}^\text{cmp} > \mathcal{E}^\text{our}$ & 2873 & 74.0\% & (-3.3, 2.4) & (0.016, 0.023) & 74.6 \\
$N_c^\text{cmp} < N_c^\text{our} \wedge \mathcal{E}^\text{cmp} \le \mathcal{E}^\text{our}$ & 3 & 0.1\% & (-44.7, 56.2) & (-0.551, 0.776) & 3.0 \\
$\text{Total}$ & 3885 & 100.0\% & (1.8, 120.4) & (0.009, 0.277) & 59.6 \\
\bottomrule
\end{tabular}
\label{tab:zz}
}
\end{table}


\begin{table}[t]
\centering
\caption{Comparison of quadrilateral meshes generated using different parameterization methods, all followed by libQEx for quadrangulation. \#F denotes the number of faces in the resulting quadrilateral mesh, and \#S denotes the number of singularities. 
IA and EL represent the interior angles and edge lengths, respectively, where each pair of values in brackets indicates the mean and standard deviation. 
The last column reports the total runtime in seconds.}
\begin{tabular}{cccccc}
\toprule
Method &
\#F &
\#S &
IA  &
EL &
Time (s)\\
\midrule
IOQ + libigl & 3520 & 25 & (89.3, 8.1) & (3.4, 0.4) & 162.1 \\
\cite{zhang2023practical} + BFF & 3617 & 14 & (89.3, 7.1) & (3.3, 0.6) & 67.2 \\
ours + BFF & 3222 & 11 & (89.2, 6.3) & (3.5, 0.6) & 3.6\\
\bottomrule
\end{tabular}
\label{tab:quad_eval}
\end{table}

\section*{Acknowledgments}\label{sec:acknowledgments}
We would like to thank the anonymous reviewers for their constructive suggestions and comments.
This work is supported by the Provincial Natural Science Foundation of Anhui, PR China (2408085QF197), the Fundamental Research Funds for the Central Universities and the National Natural Science Foundation of China (62272429, 62025207, 92570201).

\ifCLASSOPTIONcaptionsoff
  \newpage
\fi

\bibliographystyle{IEEEtran}
\bibliography{src/reference}


\begin{IEEEbiography}[{\includegraphics[width=1in,height=1.25in,clip,keepaspectratio]{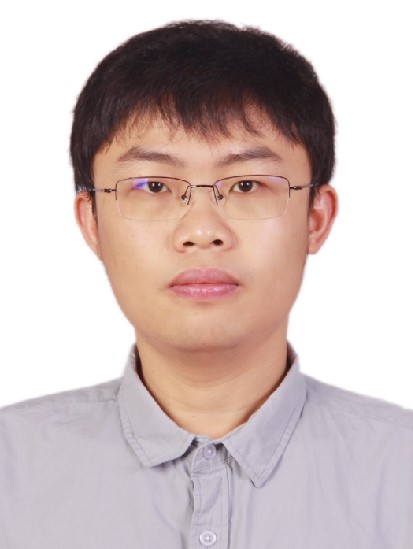}}]{Wei Du}
received a BSc degree in 2020 from the University of Science and Technology of China. He is currently a PhD candidate at the School of Mathematical Sciences, University of Science and Technology of China. His research interests include mesh generation and geometric processing. His research work can be found at his research website: \url{https://yanyiss.github.io/}.
\end{IEEEbiography}
\begin{IEEEbiography}[{\includegraphics[width=1in,height=1.25in,clip,keepaspectratio]{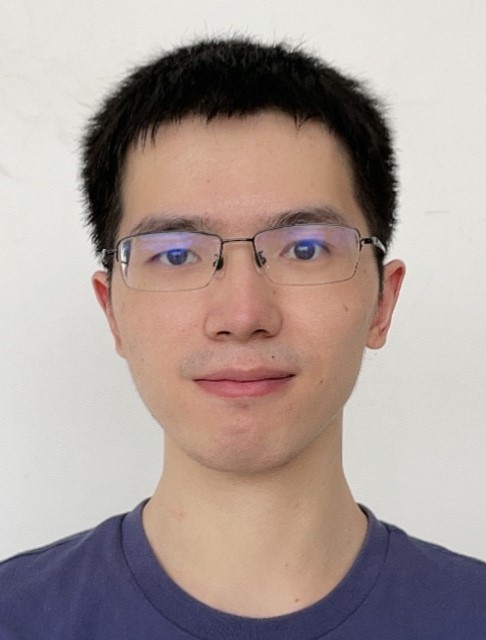}}]{Qing Fang}
received his BSc degree in 2015 and his PhD degree in 2021 from the University of Science and Technology of China (USTC), where he is currently an assistant researcher in the School of Mathematical Sciences. His research interests include geometric processing and modeling. The research work can be found at the website: \url{https://qingfang1208.github.io/}.
\end{IEEEbiography}
\begin{IEEEbiography}[{\includegraphics[width=1in,height=1.25in,clip,keepaspectratio]{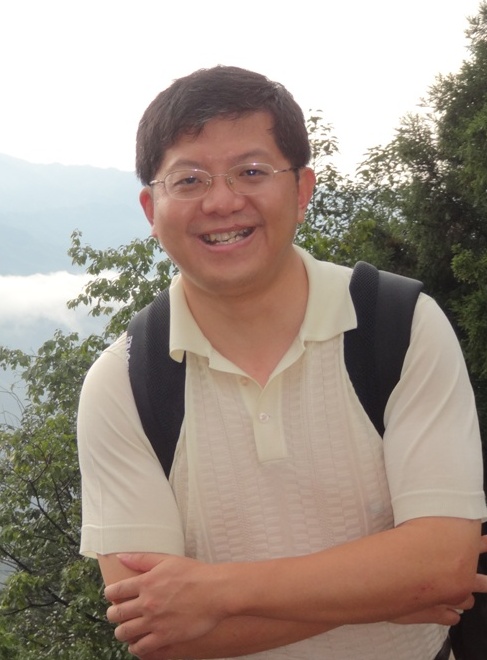}}]{Ligang Liu}
is a Professor at the School of Mathematical Sciences, University of Science and Technology of China. His research interests include computer graphics and CAD/CAE. His work on light-weight designing for fabrication at Siggraph 2013 was awarded as the first Test-of-Time Award at Siggraph 2023. \url{http://staff.ustc.edu.cn/~lgliu}.
\end{IEEEbiography}
\begin{IEEEbiography}[{\includegraphics[width=1in,height=1.25in,clip,keepaspectratio]{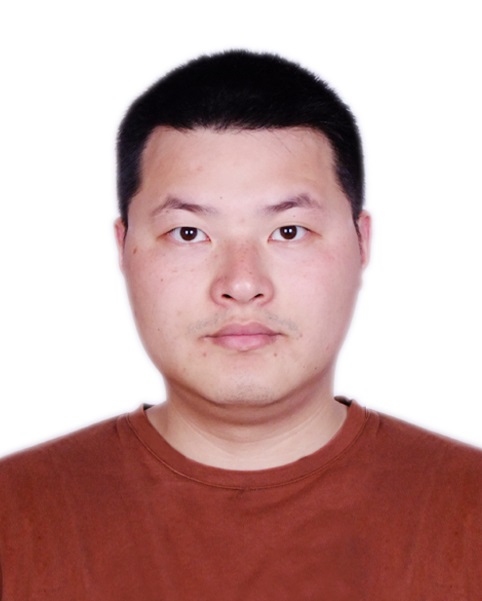}}]{Xiao-Ming Fu}
received a BSc degree in 2011 and a Ph.D. in 2016 from the University of Science and Technology of China. He is an associate professor at the School of Mathematical Sciences, University of Science and Technology of China. His research interests include geometric processing and computer-aided geometric design. His research work can be found at his website: \url{http://staff.ustc.edu.cn/~fuxm/}.
\end{IEEEbiography}

\end{document}


\title{Efficient Computation of Integer-constrained Cones for Conformal Parameterizations - Supplements}

\author{Wei~Du, Qing~Fang, Ligang~Liu, Xiao-Ming~Fu
\IEEEcompsocitemizethanks
{
\IEEEcompsocthanksitem W. Du, Q. Fang, L. Liu, and X. Fu are with the School of Mathematical Sciences, University of Science and Technology of China. 
\protect\\
E-mail: duweiyou@mail.ustc.edu.cn, fq1208@mail.ustc.edu.cn, lgliu@ustc.edu.cn, fuxm@ustc.edu.cn.
\IEEEcompsocthanksitem Corresponding author: Qing~Fang.
}
}

\markboth{Journal of \LaTeX\ Class Files,~Vol.~14, No.~8, December~2019}%
{Shell \MakeLowercase{\textit{et al.}}: Bare Demo of IEEEtran.cls for Computer Society Journals}

\maketitle
\IEEEdisplaynontitleabstractindextext
\IEEEpeerreviewmaketitle

\section{Computation on Lagrangian multiplier \(h\)}\label{discussion_on_h}

According to \cite{li2023efficient}, the boundary condition for Yamabe equation is set as \(B(u)=0\). For the cones, additional boundary conditions are required as 
\begin{equation}
    u=\phi\quad\text{on}\quad\bigcup_{c\in\mathcal{C}} \partial\delta_c,
\end{equation}
where \(\phi\) is a known function with respect to \(k^\text{ori}\) and angles of cones. If cones \(\{c_i\}_{i=1}^{N_c}\) move on the mesh to decrease the area distortion, the formulation is
\begin{equation}\label{objective}
    \begin{aligned}
        \min \quad& \mathcal{E}\\
        s.t.\quad &\Delta u=k^\text{ori}\quad\text{in}\quad \mathcal{M}\backslash\bigcup_{c\in\mathcal{C}}\delta_c,\\
        &B(u)=0\quad \text{on}\quad \partial \mathcal{M},\\
        &u=\phi\quad\text{on}\quad\bigcup_{c\in\mathcal{C}} \partial\delta_c.
    \end{aligned}
\end{equation}
Therefore, the associated Lagrangian function of~\eqref{objective} is
\begin{equation}
    \begin{aligned}
        \mathcal{L}(u,h,\lambda)&=\mathcal{E}+\int_{\mathcal{M}\backslash\bigcup_{c\in\mathcal{C}} \delta_c}h(\Delta u-k^\text{ori})dA\\
        &+\sum_{c\in \mathcal{C}}\int_{\partial\delta_c}\lambda_{\delta}(u-\phi)ds+\int_{\partial \mathcal{M}}\lambda_{\mathcal{M}}B(u)ds.
    \end{aligned}
\end{equation}
By taking variations of the Lagrangian with respect to \(u,h,\lambda_\delta,\lambda_\mathcal{M}\) and simplifying with the Green's formula, the following equations can be obtained:
\begin{equation}
    \begin{aligned}
        \Delta h&=-2u\quad\text{in}\quad\mathcal{M}\backslash\bigcup_{c\in\mathcal{C}}\delta_c,\\
        h&=0\quad\text{on}\quad\bigcup_{c\in\mathcal{C}} \partial\delta_c,\\
        \lambda_\delta&=-\frac{\partial h}{\partial n}\quad\text{on}\quad\bigcup_{c\in\mathcal{C}} \partial\delta_c,\\
        \lambda_\mathcal{M}&=0\quad\text{on}\quad\partial\mathcal{M}.
    \end{aligned}
\end{equation}

As a result, according to \cite{sharp2018variational}, the normal directional derivative is
\begin{equation}\label{eq:gradient}
    \begin{aligned}
        \nabla_n\mathcal{E}&=\frac{1}{2\mathcal{E}}(u^2+\lambda_\delta\frac{\partial u}{\partial n})\\
        &=\frac{1}{2\mathcal{E}}(u^2-\frac{\partial u}{\partial n}\frac{\partial h}{\partial n})
    \end{aligned}
\end{equation}

In our case, the cone angles remain fixed during movement. To calculate the directional derivative with respect to the moving direction $n$ of any cone $c_i\in\mathcal{C}$, we consider the scenario where only $c_i$ is optimized, while the other cones remain fixed. By the Gauss-Bonnet theorem, the cone angle of $c_i$ always remains unchanged. Thus, the formulation can be  
\begin{equation}\label{newobjective}
    \begin{aligned}
        \min \quad& \mathcal{E}\\
        s.t.\quad &\Delta u=k^\text{ori}-k^\text{tar}\quad\text{in}\quad \mathcal{M}\backslash\delta_{c_i},\\
        &B(u)=0\quad \text{on}\quad \partial \mathcal{M},\\
        &u=\phi\quad\text{on}\quad\partial\delta_{c_i},
    \end{aligned}
\end{equation}
where $k^\text{tar}$ is zero except at cones $\{c_j\in\mathcal{C},j\neq i\}$ where it is a Dirac measure. The associated Lagrangian function of~\eqref{newobjective} is
\begin{equation}
    \begin{aligned}
        \mathcal{L}(u,h,\lambda)&=\mathcal{E}+\int_{\mathcal{M}\backslash\delta_{c_i}}h(\Delta u-k^\text{ori}+k^\text{tar})dA\\
        &+\int_{\partial\delta_{c_i}}\lambda_\delta(u-\phi)ds+\int_{\partial \mathcal{M}}\lambda_{\mathcal{M}}B(u)ds.
    \end{aligned}
\end{equation}
Using variations, we can obtain:
\begin{equation}\label{eq:newvariation}
    \begin{aligned}
        \Delta h&=-2u\quad\text{in}\quad\mathcal{M}\backslash\delta_{c_i},\\
        h&=0\quad\text{on}\quad\partial\delta_{c_i},\\
        \lambda_\delta&=-\frac{\partial h}{\partial n}\quad\text{on}\quad\partial\delta_{c_i},\\
        \lambda_\mathcal{M}&=0\quad\text{on}\quad\partial\mathcal{M}.
    \end{aligned}
\end{equation}
Then we can use the~\eqref{eq:gradient} to get the directional derivative with respect
to the moving direction $n$ of the cone $c_i$. Let $\delta_{c_i}$ is a small neighborhood of $c_i$ with radius $\varepsilon$. Since $\varepsilon$ can be chosen to be arbitrarily small, the $h$ function in~\eqref{eq:newvariation} is a particular solution to the following PDE:
\begin{equation}\label{eq:generalpde}
    \Delta h=-2u\quad\text{in}\quad\mathcal{M},
\end{equation}
where $h(c_i)=0$. Since the $\Delta$ operator has a constant eigenvalue function, the term $\frac{\partial h}{\partial n}$ is the same for any general solution of~\eqref{eq:generalpde}.

\section{Surfaces with boundaries}\label{section:surfaces_with_boundaries}
\paragraph{Genus-zero surfaces with boundaries}
For meshes with boundaries, the Yamabe equation comes with boundary conditions~\cite{aubin2013some}. 
If Dirichlet boundary condition \(\mathbf{u}_\mathbf{b}=\mathbf{b}\) is applied, the discretization of Yamabe equation becomes:
\begin{equation}\label{yamabe_with_dirichlet}
\mathbf{L}_\text{II}\mathbf{u}_\text{I}+\mathbf{L}_\text{IB}\mathbf{b}=\mathbf{k}_\text{I}^\text{tar}-\mathbf{k}_\text{I}^\text{ori},
\end{equation}
where the subscripts \(\mathbf{I}\) and \(\mathbf{B}\) denote the index set of interior and boundary vertices, respectively.
Since the matrix \(\mathbf{L}_\text{II}\) is positive definite, it guarantees that \(\mathbf{u}_\text{I}\) is uniquely determined by the target angle \(\mathbf{k}^\text{tar}\). 
As a result, \(\mathbf{u}_\text{I}\) can be written as  
\begin{equation}
    \mathbf{u}_\text{I} = \mathbf{L}_\text{II}^{-1} \left( \frac{\pi}{2} \mathbf{T} \mathbf{z}^\text{int} - \mathbf{k}_\text{I}^\text{ori} - \mathbf{L}_\text{IB} \mathbf{b} \right).
\end{equation}

If the Neumann boundary conditions is given, the discrete Yamabe equation becomes:
\begin{equation}\label{yamabe_with_neumann}
\begin{pmatrix}
\mathbf{L}_\text{II}&\mathbf{L}_\text{IB}\\
\mathbf{L}_\text{BI}&\mathbf{L}_\text{BB}
\end{pmatrix}
\begin{pmatrix}
    \mathbf{u}_\text{I}\\
    \mathbf{u}_\text{B}
\end{pmatrix}
+
\begin{pmatrix}
\mathbf{0}\\
\mathbf{h}
\end{pmatrix}
=
\begin{pmatrix}
    \mathbf{k}_\text{I}^\text{tar}-\mathbf{k}_\text{I}^\text{ori}\\
    \mathbf{k}_\text{B}^\text{tar}-\mathbf{k}_\text{B}^\text{ori}
\end{pmatrix},
\end{equation}
where \(\mathbf{h}\) denotes the directional derivative of the log conformal factor along the boundary normal.
The matrix on the left-hand side of \eqref{yamabe_with_neumann} is rank-deficient by one, so the approach described in Sec. III can be adopted to handle this case.

Therefore, regardless of whether Dirichlet or Neumann boundary conditions are applied, \(\mathbf{u}\) can still be expressed in terms of \(\mathbf{z}^\text{int}\), allowing us to derive a reduced formulation. 
All techniques in Sec. III remain practical in these settings except that the topology constraint is removed, i.e., we do not enforce \(\sum_{i=1}^{N_c}z_{c_i}\) to be a constant value.

\paragraph{Nonzero genus surfaces with boundaries}
For nonzero genus meshes with boundaries, the difference in treatment compared to boundary-free meshes is the same as in the genus-zero case: we impose boundary conditions and remove the topological constraints. If the Dirichlet boundary conditions are applied, we similarly do not modify the Laplacian matrix.

\begin{figure}[t]
  \centering
  \begin{overpic}[width=0.99\linewidth]{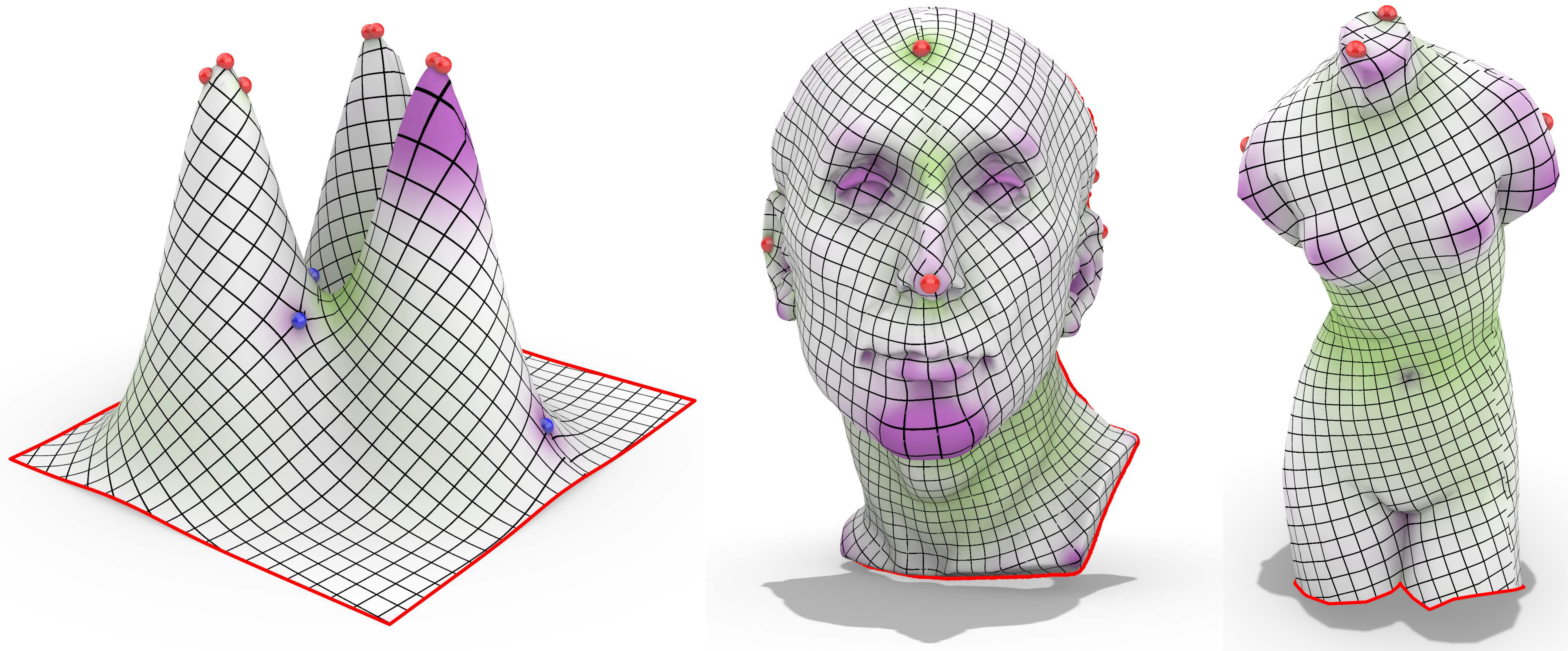}
    {
 \put(12,-3){\small (14, 0.158, 1.7)}
 \put(50,-3){\small (5, 0.161, 3.8)}
 \put(79,-3){\small (4, 0.159, 0.8)}
    }
  \end{overpic}
  \vspace{1mm}
  \caption{
  Three surfaces with boundaries.
  }
  \label{fig:surfaces_with_boundary}
\end{figure}

Fig.~\ref{fig:surfaces_with_boundary} shows three models with boundaries handled by our method, where the Gauss-Bonnet constraint is no longer imposed, allowing positive and negative cones to be placed freely.


\tb{
\section{The rank of \(\mathbf{L}_\text{g}\)}
If, at each intersection, we use only a single variable instead of the three variables \(u_3,u_6,u_9\) as in Fig. 9, then 
\begin{equation}
\mathbf{L}_\text{g}=
\begin{pmatrix}
    A&B\\
    C&D
\end{pmatrix},
\end{equation}
with \(A\) being the Laplacian matrix and \(C=D=\mathbf{0}\).
In this case, \(\operatorname{rank}(\mathbf{L}_g)=N-1\). However, when we introduce two variables at each intersection, two columns in \(C\) and \(D\) become nonzero, which increases the rank by 2. Since there are \(g\) intersections in total, the rank of \(\mathbf{L}_g\) becomes \(N+2g-1\).}

\ifCLASSOPTIONcaptionsoff
  \newpage
\fi

\bibliographystyle{IEEEtran}
\bibliography{src/reference}